# DEVS/SOA: A Cross-Platform Framework for Net-centric Modeling & Simulation in DEVS Unified Process

Saurabh Mittal, Ph.D.
*José L. Risco-Martín, Ph.D.
Bernard P. Zeigler, Ph.D.

Arizona Center for Integrative Modeling and Simulation
Electrical and Computer Engineering,
University of Arizona
Tucson, AZ

*Departamento de Arquitectura de Computadores y Automática
Facultad de Informática
Universidad Complutense de Madrid
Madrid, Spain

*Abstract*

*Discrete EVent Specification (DEVS) environments are known to be implemented over middleware systems such as HLA, RMI, CORBA and others. DEVS exhibits concepts of systems theory and modeling and supports capturing the system behavior from the physical and behavioral perspectives. Further, they are implemented using Object-oriented languages like Java and C++. This research work uses the Java platform to implement DEVS over a Service Oriented Architecture (SOA) framework. Called the DEVS/SOA, the framework supports a development and testing environment known as DEVS Unified Process that is built on a model-continuity-based life cycle methodology. DEVS Unified Process allows DEVS-based Modeling and Simulation (M&S) over net-centric platforms using DEVS/SOA. This framework also provides the crucial feature of run-time composability of coupled systems using SOA. We describe the architecture and designs of the both the server and the client. The client application communicates with multiple servers hosting DEVS simulation services. These Simulation services are developed using the proposed symmetrical services architecture wherein the server can act as both a service provider and a service consumer contrary to the unidirectional client-server paradigm. We also discuss how this Services based architecture provides solutions for cross-platform distributed M&S. We demonstrate DEVS/SOA framework with a scenario of Joint Close Air Support specified in Business Process Modeling Notation (BPMN). We also provide a real-world application of Network health monitoring using DEVS/SOA layered architectural framework.*

## *1. Introduction*

DEVS environments such as DEVSJAVA, DEVS-C++, and others [9] are embedded in object-oriented implementations, they support the goal of representing executable model architectures in an object-oriented representational language. As a mathematical formalism, DEVS is platform independent, and its implementations adhere to the DEVS protocol so that DEVS models easily translate from one form (e.g., C++) to another (e.g., Java) [10]. Moreover, DEVS environments, such as DEVSJAVA, execute on commercial, off-the-shelf desktops or workstations and employ state-of-the-art libraries to produce graphical output that complies with industry and international standards. DEVS environments are typically open architectures that have been extended to execute on various middleware such as the DoD's HLA standard, CORBA, SOAP, and others and can be readily interfaced to other engineering and simulation and modeling tools [2, 9, 27, 28, 30]. Furthermore, DEVS operation over web middleware (SOAP) enables it to fully participate in the net-centric environment of the Global Information Grid/ Service Oriented Architecture (GIG/SOA) [8]. As a result of recent advances, DEVS can support model continuity through a simulation-based development and testing life cycle [2]. This means that the mapping of high-level requirement specifications into lower-level DEVS formalizations enables such specifications to be thoroughly tested in virtual simulation environments before being easily and consistently transitioned to operate in a real environment for further testing and fielding.

DEVS formalism categorically separates the Model, the Simulator and the Experimental frame. However, one of the major problems in this kind of mutually exclusively system is that the formalism implementation is itself limited by the underlying programming language. In other words, the model and the simulator exist in the same programming language. Consequently, legacy models as well as models that are available in one implementation are hard to translate from one language to another even though both the implementations are object oriented. Other constraints like libraries inherent in C++ and Java are another source of bottleneck that prevents such interoperability.

**Brief Overview of Capabilities Provided by DEVS**

The prime motivation comes from an editorial by Carstairs [1] that demands a M&S framework at higher levels of system specifications where System of systems interact together using net-centric platform. At this level, model interoperability is one of the major concerns. The motivation for this work stems from this need of model interoperability between the disparate simulator implementations and provides a means to make the simulator transparent to model execution. DEVS, which is known to be component-based system, based on forma systems theoretical framework is the preferred means. Table 1 outlines how it could provide solutions to the challenges in net-centric design and evaluation. The net-centric DEVS framework requires enhancement to the basic DEVS capabilities, which are provided in later sections.

| Desired M&S Capability for T&E | Solutions Provided by DEVS Technology |
|---|---|
| Support of DoDAF need for executable architectures using M&S such as mission based testing for GIG SOA | DEVS Unified Process [31] provides methodology and SOA infrastructure for integrated development and testing, extending DoDAF views [32]. |
| Interoperability and cross-platform M&S using GIG/SOA | Simulation architecture is layered to accomplish the technology migration or run different technological scenarios [13, 17]. Provide net-centric composition and integration of DEVS 'validated' models using Simulation Web Services [19] |
| Automated test generation and deployment in distributed simulation | Separate a model from the act of simulation itself, which can be executed on single or multiple distributed platforms [10]. With its bifurcated test and development process, automated test generation is integral to this methodology [18]. |
| Test artifact continuity and traceability through phases of system development | Provide rapid means of deployment using model-continuity principles and concepts like "simulation becomes the reality" [2]. |
| Real time observation and control of test environment | Provide dynamic variable-structure component modeling to enable control and reconfiguration of simulation on the fly [14-17]. Provide dynamic simulation tuning, interoperability testing and benchmarking. |

**Table 1:** Solutions provided by DEVS technology to support of M&S for T&E

Furthermore, this work aims to develop and evaluate distributed simulation using the web service technology. After the development of World Wide Web, many efforts in the distributed simulation field have been made for modeling, executing simulation and creating model libraries that can be assembled and executed over WWW. By means of XML and web services technology these efforts have entered upon a new phase. We proposed DEVS Modeling Language (DEVSML) [19] that is built on eXtensible Markup Language (XML) [29] as the preferred means to provide such transparent simulator implementation. A prototype simulation framework called DEVS/SOA has been implemented using web services technology. The central point resides in executing the simulator as a web service. The development of this kind of frameworks will help to solve large-scale problems and guarantees interoperability among different networked systems and specifically DEVS-validated models. This paper focuses on the overall approach, and the symmetrical SOA-Based architecture that allows for DEVS execution as a Simulation SOA.

The paper is organized as follows. The next section provides information about the related work in distributed simulation and DEVS standardization efforts. Section 3 describes the underlying technologies such as DEVS, Web Services, XML, and DEVS Modeling Language (DEVSML). Section 4 introduces DEVS/SOA and presents its relationship with DEVS Unified Process (DUNIP) along with DEVSML. It also compares with Model Driven Architecture (MDA) with DUNIP. Section 5 presents the DEVS/SOA distributed simulation framework in detail. It provides the symmetrical web services architecture, the conceptual design, the implemented packages and the Web

Service design architecture. By symmetrical server we mean that it acts as both a service provider and a service consumer. This section presents both the server and client designs. Section 6 extends the DEVS/SOA framework towards cross-platform distributed simulation framework and provides theoretical basis to conduct cross-platform simulation on SOA. It discusses Interoperability. Section 7 provides one illustrative example that describes the complete life-cycle in DEVS Unified Process and how a model is made Net-centric executable using DEVS/SOA. It also provide two other applications that relate to Mission Thread modeling as applicable to DoDAF and a proactive network health monitoring system. Finally, Section 8 provides conclusions and open research lines.

## 2. Related Work

There have been a lot of efforts in the area of distributed simulation using parallelized DEVS formalism. Issues like 'causal dependency' [10] and 'synchronization problem' [20] have been adequately dealt with solutions like: 1. restriction of global simulation clock until all the models are in sync, or 2. rolling back the simulation of the model that has resulted in the causality error. Our chosen method of web centric simulation does not address these problems as they fall in a different domain. In our proposed work, the simulation engine rests solely on the Server. Consequently, the coordinator and the model simulators are always in sync.

Most of the existing web-centric simulation efforts consist of the following components:
1. *the Application*: The top level coupled model with (optional) integrated visualization.
2. *Model partitioner*: Element that partitions the model into various smaller coupled models to be executed at a different remote location
3. *Model deployer*: Element that deployed the smaller partitioned models to different locations
4. *Model initializer*: Element that initializes the partitioned model and make it ready for simulation
5. *Model Simulator*: Element that coordinate with root coordinator about the execution of partitioned model execution.

The Model Simulator design is almost same in all of the implementation and is derived directly from parallel DEVS formalism [10]. There are however, different methods to implement the former four elements. DEVS/Grid [21] uses all the components above. DEVS/P2P [22] implements step 2 using hierarchical model partitioning based on cost-based metric. DEVS/RMI [30] has a configuring engine that integrates the functionality of step 1, 2 and 3 above. DEVS/Cluster [23] is a multi-threaded distributed DEVS simulator built on CORBA, which again, is focused towards development of simulation engine.

As stated earlier, the efforts have been in the area of using the parallel DEVS and implementing the simulator engine in the same language as that of the model.

These efforts are in no means similar to what we had proposed in our paper [19]. Our work is focused towards interoperability at the application level, specifically, at the model level and hiding the simulator engine as a whole. We are focused towards taking XML just as a communication middleware, as used in SOAP, for existing DEVS models, but not as complete solution in itself. We would like the user or designer to code the behavior in any of the programming languages and let the DEVSML SOA architecture be responsible to create a coupled model, integrating code in either of the languages and delivering us with an executable model that can be simulated. The user need not learn any new syntax, any new language; however, what he must use is the standardized version of P-DEVS implementation such as DEVSJAVA Version 3.0 [9] (maintained at www.acims.arizona.edu).

This kind of capability where the user can integrate his model from models stored in any web repository, whether it contained public models of legacy systems or proprietary standardized models will provide more benefit to the industry as well as to the user, thereby truly realizing the model-reuse paradigm.

In further sections we will provide details about the DEVS/SOA server and client, design of DEVS Simulator interface and standardized libraries that are used in our implementation.

## *3. Underlying Technologies*

## **3.1 DEVS**

DEVS formalism consists of models, the simulator and the Experimental Frame. We will focus our attention to the two types of models i.e. atomic and coupled models. The atomic model is the irreducible model definitions that specify the behavior for any modeled entity. The coupled model is the aggregation/composition of two or more atomic models connected by explicit couplings. The coupled model N can itself be a part of component in a larger coupled model system giving rise to a hierarchical DEVS model construction. Detailed descriptions about DEVS Simulator, Experimental Frame and of both atomic and coupled models can be found in [10]. Next we review some of the background required for discussion on the usage of DEVS foundation.

### **3.1.1 DEVS Specification**

The DEVS formalism was introduced by Bernard Zeigler [10] to provide a mean of modeling discrete event systems in a hierarchical and modular way. DEVS exhibits the concepts of system theory and modeling, and supports capturing the system behavior in the physical and behavioral perspectives. A DEVS model can be either an atomic or coupled model. In the DEVS formalism, a large system can be modeled by both atomic and coupled models. The atomic model is the basic model that describes the behavior of a component. A Discrete Event System specification (DEVS) atomic model is defined by the structure in Figure 1.

$M = <X, S, Y, \delta_{int}, \delta_{ext}, \lambda, ta>$

where

$X$ is the set of input values
$S$ is the set of state
$Y$ is the set of output values
$\delta_{int}: S \to S$ is the internal transition function
$\delta_{ext}: Q \times X \to S$ is the external transition function, where
$Q = \{(s,e) | s \in S, 0 \leq e \leq ta(s)\}$ is the total state set, and
e is the time elapsed since last transition
$\lambda: S \to Y$ is the output function
$ta: S \to R_{0,inf}^{+}$ is the time advance function

**Figure 1:** Classic DEVS Specification

Atomic and coupled models can be simulated using sequential computation or various forms of parallelism. The basic parallel DEVS formalism extends the classic DEVS by allowing bags of inputs to the external transition function, and it introduces the confluent transition function to control the collision behavior when receiving external events at the time of the internal transition. The parallel DEVS atomic model is defined by the structure in Figure 2.

$M = <X, S, Y, \delta_{int}, \delta_{ext}, \delta_{con}, \lambda, ta>$

where

$X$ is the set of input values
$S$ is the set of state
$Y$ is the set of output values
$\delta_{int}: S \to S$ is the internal transition function
$\delta_{ext}: Q \times X^b \to S$ is the external transition function,
where $X^b$ is a set of bags over elements in X, Q is the total state set.
$\delta_{con}: S \times X^b \to S$ is the confluent transition function,
subject to $\delta_{con}(s, \Phi) = \delta_{int}(s)$
$\lambda: S \to Y^b$ is the output function
$ta: S \to R_{0,inf}^{+}$ is the time advance function

**Figure 2:** Parallel DEVS Specification

A DEVS-coupled model designates how atomic models can be coupled together and how they interact with each other to form a complex model. The coupled model can be employed as a component in a larger coupled model and can construct complex models in a hierarchical way. The specification provides component and coupling information. The coupled DEVS model is defined as the structure in Figure 3.

$$M = <X, Y, D, \{M_{ij}\}, \{I_j\}, \{Z_{ij}\}>$$

Where

    **X** is a set of inputs
    **Y** is a set of outputs
    **D** is a set of DEVS component names
    For each i ∈ D,
        $M_i$ is a DEVS component model
        $I_i$ is the set of influences for I
    For each j ∈ $I_i$,
        $Z_{ij}$ is the i-to-j output translation function.

**Figure 3:** Coupled DEVS Specification

Three different DEVS formalisms have been introduced. The classic DEVS formalism treats components sequentially, and the parallel DEVS formalism treats components concurrently. These formalisms also include the means to build coupled model from atomic models.

### 3.1.2 Hierarchy of Systems specifications

Systems theory deals with a hierarchy of system specifications which defines levels at which a system may be known or specified. Table 2 shows this Hierarchy of System Specifications (in simplified form, see [10]).

- At level 0 we deal with the input and output interface of a system.
- At level 1 we deal with purely observational recordings of the behavior of a system. This is an I/O relation which consists of a set of pairs of input behaviors and associated output behaviors.
- At level 2 we have knowledge of the initial state when the input is applied. This allows partitioning the input/output pairs of level 1 into non-overlapping subsets, each subset associated with a different starting state.
- At level 3 the system is described by state space and state transition functions. The transition function describes the state-to-state transitions caused by the inputs and the outputs generated thereupon.
- At level 4 a system is specified by a set of components and a coupling structure. The components are systems on their own with their own state set and state transition functions. A coupling structure defines how those interact. A property of coupled system which is called "closure under coupling" guarantees that a coupled system at level 3 itself specifies a system. This property allows hierarchical construction of systems, i.e., that coupled systems can be used as components in larger coupled systems.

| Level | Name | What we specify at this level |
|---|---|---|
| 4 | Coupled Systems | System built up by several component systems which are coupled together |
| 3 | I/O System | System with state and state transitions to generate the behavior |
| 2 | I/O Function | Collection of input/output pairs constituting the allowed behavior partitioned according to the initial state the system is in when the input is applied |
| 1 | I/O Behavior | Collection of input/output pairs constituting the allowed behavior of the system from an external Black Box view |
| 0 | I/O Frame | Input and output variables and ports together with allowed values |

**Table 2**: Hierarchy of System Specifications

As we shall see in a moment, the system specification hierarchy provides a mathematical underpinning to define a framework for modeling and simulation. Each of the entities (e.g., real world, model, simulation, and experimental frame) will be described as a system known or specified at some level of specification. The essence of modeling and simulation lies in establishing relations between pairs of system descriptions. These relations pertain to the validity of a system description at one level of specification relative to another system description at a different (higher, lower, or equal) level of specification.

Based on the arrangement of system levels as shown in Table 2, we distinguish between vertical and horizontal relations. A vertical relation is called an association mapping and takes a system at one level of specification and generates its counterpart at another level of specification. The downward motion in the structure-to-behavior direction, formally represents the process by which the behavior of a model is generated. This is relevant in simulation and testing when the model generates the behavior which then can be compared with the desired behavior.

The opposite upward mapping relates a system description at a lower level with one at a higher level of specification. While the downward association of specifications is straightforward, the upward association is much less so. This is because in the upward direction information is introduced while in the downward direction information is reduced. Many structures exhibit the same behavior and recovering a unique structure from a given behavior is not possible. The upward direction, however, is fundamental in the design process where a structure (system at level 3) has to be found which is capable to generate the desired behavior (system at Level 1).

### 3.1.3 Framework for Modeling & Simulation

The *Framework for M&S* as described in [10], establishes *entities* and their *relationships* that are central to the M&S enterprise (see Figure 2). The entities of the framework are *source system, experimental frame, model,* and *simulator;* they are linked by the *modeling* and the *simulation* relationships. Each entity is formally characterized as a system at an appropriate level of specification within a generic dynamic system. See [10] for detailed discussion.

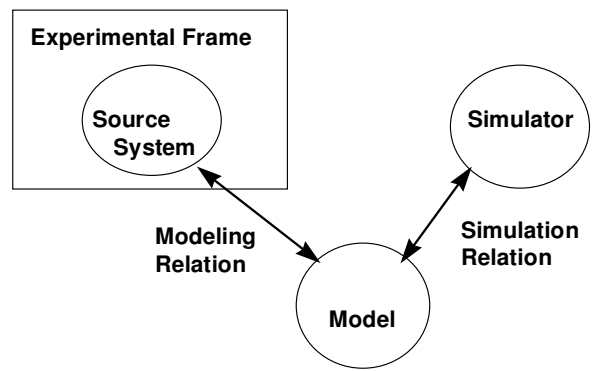

**Figure 4:** Framework Entities and Relationships

### 3.1.4 Model Continuity

Model continuity refers to the ability to transition as much as possible of a model specification through the stages of a development process. This is opposite to the discontinuity problem where artifacts of different design stages are disjointed and thus cannot be effectively consumed by each other. This discontinuity between the artifacts of different design stages is a common deficiency of most design methods and results in inherent inconsistency among analysis, design, test, and implementation artifacts [11]. Model continuity allows component models of a distributed real-time system to be tested incrementally, and then deployed to a distributed environment for execution. It supports a design and test process having 4 steps (see [11]),
   1) Conventional simulation to analyze the system under test within a model of the environment linked by abstract sensor/actuator interfaces.

2) Real-time simulation, in which simulators are replaced by a real-time execution engines while leaving the models unchanged.
3) Hardware-in-the-loop (HIL) simulation in which the environment model is simulated by a DEVS real-time simulator on one computer while the model under test is executed by a DEVS real-time execution engine on the real hardware.
4) Real execution, in which DEVS models interact with the real environment through the earlier established sensor/actuator interfaces that have been appropriately instantiated under DEVS real-time execution.

Model continuity reduces the occurrence of design discrepancies along the development process, thus increasing the confidence that the final system realizes the specification as desired. Furthermore, it makes the design process easier to manage since continuity between models of different design stages is retained.

## 3.2 Web Services and Interoperability using XML

Service oriented Architecture (SOA) framework is a framework consisting of various W3C standards, in which various computational components are made available as 'services' interacting in an automated manner towards achieving machine-to-machine interoperable interaction over the network. The interface is specified using Web Service Description language (WSDL) [25] that contains information about ports, message types, port types, and other relating information for binding two interactions. It is essentially a client server framework, wherein client request a 'service' using SOAP message that is transmitted via HTTP in XML format. A Web service is published by any commercial vendor at a specific URL to be consumed/requested by another commercial application on the Internet. It is designed specifically for machine-to-machine interaction. Both the client and the server encapsulate their message in a SOAP wrapper.

## 3.3 DEVSML

DEVSML is a way of representing DEVS models in XML language. This DEVSML is built on JAVAML [7], which is XML implementation of JAVA. The current development effort of DEVSML takes its power from the underlying JAVAML that is needed to specify the 'behavior' logic of atomic and coupled models. The DEVSML models are transformable back'n forth to java and to DEVSML. It is an attempt to provide interoperability between various models and create dynamic scenarios. The layered architecture of the said capability is shown in Figure 5.

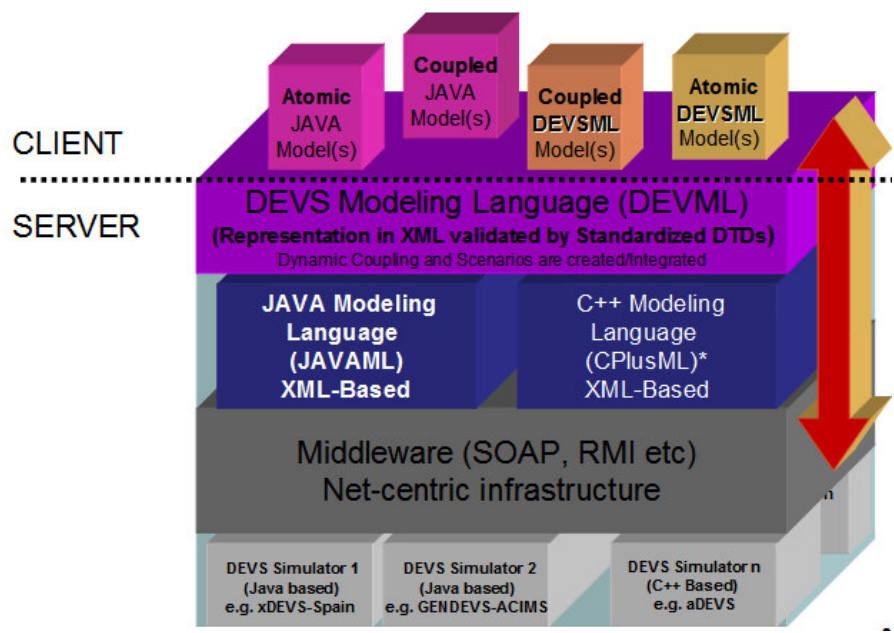

**Figure 5:** DEVS Transparency and Net-centric model interoperability using DEVSML. Client and Server categorization is done for DEVS/SOA implementation

At the top is the application layer that contains model in DEVS/JAVA or DEVSML. The second layer is the DEVSML layer itself that provides seamless integration, composition and dynamic scenario construction resulting in portable models in DEVSML that are complete in every respect. These DEVSML models can be ported to any remote location using the net-centric infrastructure and be executed at any remote location. Another major advantage of such capability is total simulator 'transparency'. The simulation engine is totally transparent to model execution over the net-centric infrastructure. The DEVSML model description files in XML contains meta-data information about its compliance with various simulation 'builds' or versions to provide true interoperability between various simulator engine implementations. This has been achieved for at least two independent simulation engines as they have an underlying DEVS protocol to adhere to. This has been made possible with the implementation of a single atomic DTD and a single coupled DTD that validates the DEVSML descriptions generated from these two implementations. Such run-time interoperability provides great advantage when models from different repositories are used to compose bigger coupled models using DEVSML seamless integration capabilities. More details about the implementation can be seen at [MIT07e]

## *4. Overarching DEVS Unified Process*

This section describes the refined bifurcated Model-Continuity process and how various elements like automated DEVS model generation, automated test-model generation (and net-centric simulation over SOA are put together in the process, resulting in DEVS Unified Process (DUNIP) [31]. The DEVS Unified Process (DUNIP) is built on the bifurcated Model-continuity based life-cycle methodology. The design of simulation-test framework occurs in parallel with the simulation-model of the system under design. The DUNIP process consists of the following elements:
- Automated DEVS Model Generation from various requirement specification formats
- Collaborative model development using DEVS Modeling Language (DEVSML)
- Automated Generation of Test-suite from DEVS simulation model
- Net-centric execution of model as well as test-suite over SOA

Considerable amount of effort has been spent in analyzing various forms of requirement specifications, viz, state-based, Natural Language based, Rule-based, BPMN/BPEL-based and DoDAF-based, and the automated processes which each one should employ to deliver DEVS hierarchical models and DEVS state machines [31]. Simulation execution today is more than just model execution on a single machine. With Grid applications and collaborative computing the norm in industry as well as in scientific community, a net-centric platform using XML as middleware results in an infrastructure that supports distributed collaboration and model reuse. The infrastructure provides for a platform-free specification language DEVS Modeling Language (DEVSML) [MIT07e] and its net-centric execution using Service-Oriented Architecture called DEVS/SOA [31,33]. Both the DEVSML and DEVSV/SOA provide novel approaches to integrate, collaborate and remotely execute models on SOA. This infrastructure supports automated procedures is the area of test-case generation leading to test-models. Using XML as the system specifications in rule-based format, a tool known as Automated Test Case Generator (ATC-Gen) was developed which facilitated the automated development of test models[6,18]. The integration of DEVSML and DEVS/SOA is performed with the layout as shown below in Figure 6.

Various model specification formalisms are supported and mapped into DEVSML models including UML state charts [5], a table driven state-based approach[31], Business Process Modeling Notation (BPMN) [34.35] or DoDAF-based[32]. A translated DEVSML model is fed to the DEVSML client that coordinates with the DEVSML server farm. Once the client has DEVSJAVA models, a DEVSML server can be used to integrate the client's model with models that are available at other sites to get an enhanced integrated DEVSML file that can produce a coupled DEVSML model. The DEVS/SOA enabled server can use this integrated DEVSML file to deploy the component models to assigned DEVS web-server simulated engines. The result is a distributed simulation, or alternatively, a real-time distributed execution of the coupled model.

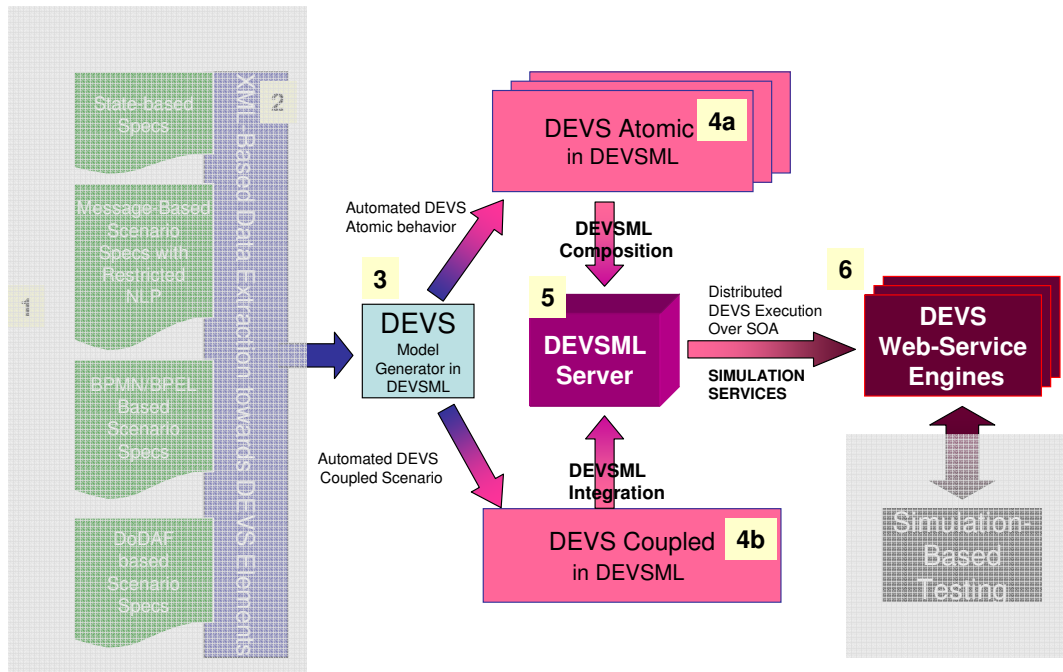

**Figure 6:** Net-centric collaboration and execution using DEVSML and DEVS/SOA

## 4.1 MDA and DUNIP

DUNIP is built on the paradigm of Model-Based Engineering, or Model-Driven Architecture (MDA). However, the scope of DUNIP goes beyond the MDA objectives. Potential concerns with the current MDA state of art include:
- MDA approach is underpinned by a variety of technical standards, some of which are yet to be specified (e.g. executable UML)
- Tools developed my many vendors are not interoperable
- MDA approach is considered too-idealistic lacking iterative nature of Software Engineering process
- MDA practice requires skilled practitioners and design requires engineering discipline not commonly available to code developers.

Further, MDA does not have any underlying Systems theory and groups like INCOSE[1] are working with OMG to adapt UML to systems engineering. Various other effort like Wegmann [3] have recommended MDA to be utilized using an underlying common systems modeling ontology. Testing is included only as an extension of UML, known as executable UML [Mel02], for which there is no current standard. Consequently, there is no testing framework that binds executable UML and simulation-based testing. Despite these shortcomings, MDA has been adopted by Joint Single Integrated Air Picture (SIAP) Systems Engineering Organization (JSSEO) and various recommendations have come forth to enhance the MDA process. JSSEO is applying MDA approach toward development of aerospace Command and Control (C2) capabilities, for which a single integrated air picture is foundational. The data-driven nature of C2 System of Systems (SoS) means that powerful MDA concepts adapt well to collaborative SoS challenges.

Current DoD enterprise-level approaches for managing SoS interoperability, like the Net Centric Operations and Warfare Reference Model (NCOW/RM), DoD Architecture Framework (DoDAF) and the Joint Technical Architecture (JTA), simply do not have the technical strength to deal with the extremely complex engineering challenges. We proposed enhanced DoDAF [32] to provide DEVS-based Model engineering. MDA as implemented by industry and adapted by JSSEO, does have the requisite technical power, but requires innovative engineering practices.

---

[1] International Council on Systems Engineering

Realizing the importance of MDA concepts and the executable profile of UML, the basic objective of which is to simulate the model, JSSEO is indirectly looking at the Modeling & Simulation domain as applicable to SoS engineering. Table 3 brings out the shortcomings of MDA in its current state and the capabilities provided by DEVS technology and in turn, DUNIP process.

| Desired M&S Capability | MDA | DUNIP |
|---|---|---|
| Need for executable architectures using M&S | Yes, although not a standard yet | Yes, underlying DEVS theory |
| Applicable to GIG SOA | Not reported yet | Yes |
| Interoperability and cross-platform M&S using GIG/SOA | -- | Yes, DEVSML and DEVSV/SOA provides cross-platform M&S using Simulation Web Services |
| Automated test generation and deployment in distributed simulation | -- | Yes, based on formal Systems theory and test-models autogeneration at various levels of System specifications |
| Test artifact continuity and traceability through phases of system development | To some extent, model becomes the application itself | Yes |
| Real time observation and control of test environment | -- | Dynamic Model Reconfiguration and run-time simulation control integral to DEVS M&S. Enhanced MVC framework is designed to provide this capability |

**Table 3:** Comparison of MDA and DUNIP

**MDA as applied to Integration of Process-Driven SOA Models**
In an independent study [36], Model Driven Software Development (MDSD) was applied to the integration of process-driven SOA models. UML2 was used as the basis towards integration. Their approach is based on the notion of domain-specific languages (DSL) for modeling various types of models. Once DSL has been identified, its meta-model is created that represents this particular modeling domain. Meta-models are defined in terms of meta-meta-model. In UML, this is the meta object facility (MOF). They created a meta-meta-model that would define both the UML2 meta-model and their selected DSL extensions. The whole objective is to find a common ground and a way to express the relationship between a meta-model and the implementation code. This kind of capability where a single meta-meta-model can be used to integrate two different DSLs towards a common model allowing specific constraints of each meta-model is very much needed in SOA domain as multiple tools and standards exist preventing such integration. To integrate two models with different DSLs, the models are first decomposed at the meta-model level, required information extracted and supplemented (on the basis of meta-meta-model), which results in an integrated model.

In our DUNIP process, such collaboration comes naturally due to the proposed DEVS atomic and coupled Document Type Definitions (DTDs) that specify any DEVS model in any domain specific language implementations. The underlying DEVS Modeling Language (DEVSML) meta-model that defines these atomic and coupled DTDs is used for validating any DEVS model. The current DEVSML implementation has successfully integrated two DSL implementations (GenDEVS-ACIMS [9] and xDEVS-Spain[4]) on common DEVSML atomic and coupled DTDs.

## *5. Distributed Simulation using DEVS/SOA*

Web-based simulation requires the convergence of simulation methodology and WWW technology (mainly Web Service technology). The fundamental concept of web services is to integrate software application as services. Web services allow the applications to communicate with other applications using open standards. We are offering DEVS-based simulators as a web service, and they must have these standard technologies: communication protocol (Simple Object Access Protocol, SOAP), service description (Web Service Description Language, WSDL), and service discovery (Universal Description Discovery and Integration, UDDI).

Figure 7 shows the framework of the proposed distributed simulation using SOA. The complete setup requires one or more servers that are capable of running DEVS Simulation Service. The capability to run the simulation service is provided by the server side design of DEVS Simulation protocol supported by the latest DEVSJAVA Version 3.1. d

The Simulation Service framework is two layered framework. The top-layer is the user coordination layer that oversees the lower layer. The lower layer is the true simulation service layer that executes the DEVS simulation protocol as a Service. The lower layer is transparent to the modeler and only the top-level is provided to the user.

The top-level has four main services:
- Upload DEVS model
- Compile DEVS model
- Simulate DEVS model (centralized)
- Simulate DEVS model (distributed)

The second lower layer provides the DEVS Simulation protocol services:
- Initialize simulator i
- Run transition in simulator i
- Run lambda function in simulator i
- Inject message to simulator i
- Get time of next event from simulator i
- Get time advance from simulator i
- Get console log from all the simulators
- Finalize simulation service

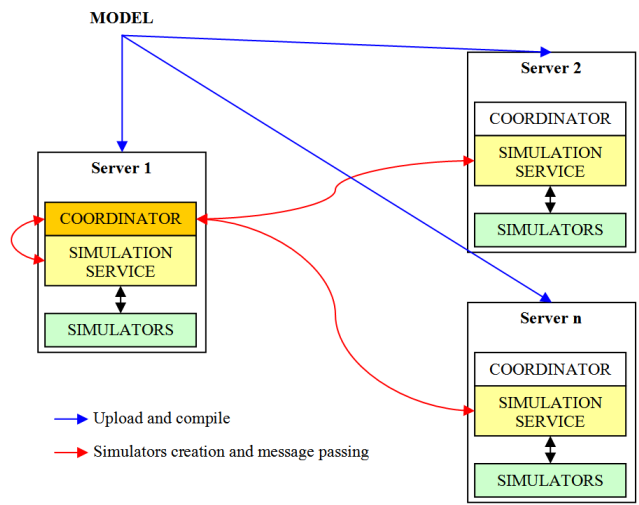

**Figure 7:** DEVS/SOA distributed architecture

The explicit transition functions, namely, the internal transition function, the external transition function, and the confluent transition function, are abstracted to a single transition function that is made available as a Service. The transition function that needs to be executed depends on the simulator implementation and is decided at the run-time. For example, if the simulator implements the Parallel DEVS (P-DEVS) formalism, it will choose among internal transition, external transition or confluent transition[2].

---
[2] The difference between P-DEVS and classic DEVS is the handling of confluent function. The DEVS/SOA framework could have been built using other simulation formalisms. In fact, our simulation services could store any

The client is provided a list of servers hosting DEVS Service. He selects some servers to distribute the simulation of his model. Then, the model is uploaded and compiled in all the servers. The main server selected creates a coordinator that creates simulators in the server where the coordinator resides and/or over the other servers selected.

Summarizing from a user's perspective, the simulation process is done through three steps (Figure 8):
1. Write a DEVS model (currently DEVSJAVA is only supported).
2. Provide a list of DEVS servers (through UDDI, for example). Since we are testing the application, these services have not been published using UDDI by now. Select N number of servers from the list available.
3. Run the simulation (upload, compile and simulate) and wait for the results.

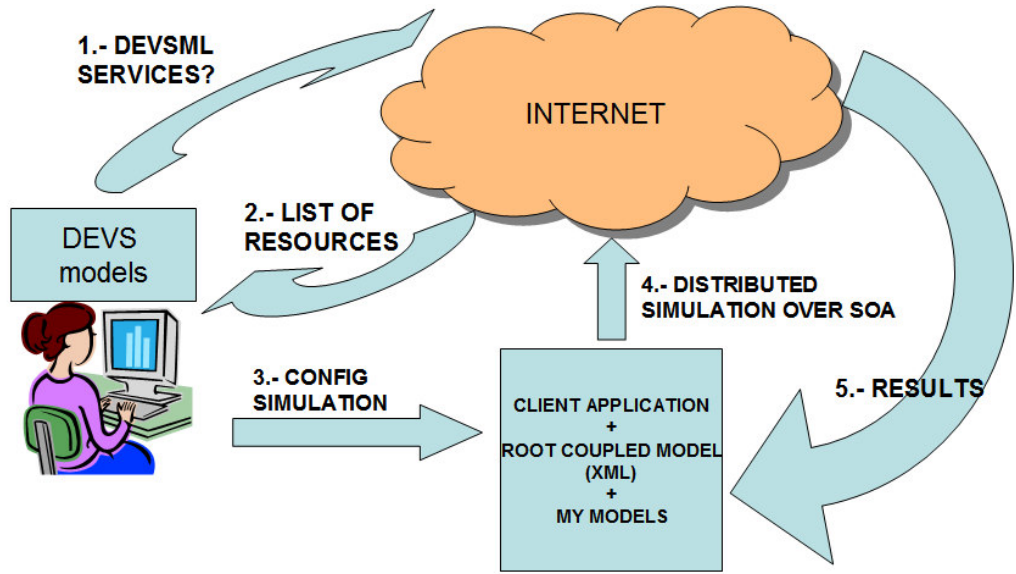

**Figure 8:** Execution of DEVS SOA-Based M&S

## 5.1 Symmetrical Services Architecture

The Web Service framework is essentially a client-server framework wherein a Server on requested by a client provides services. These services are nothing but computational code that is executed at the server's end with a valid return value. The mode of communication between the client and the server is done using standards like XML, HTTP, and SOAP. This standardized mode of communication provides interoperability between various services as the data, expressed in XML, is machine-readable.

In order to implement our DEVS/SOA framework, we have to beyond this client-server paradigm for this paradigm is not distributed in nature. Even though it operates on a Network (Internet), it is not distributed. We needed to implement a distributed framework to have the capability of distributed modeling and simulation. The distributed DEVS protocol has two types of components i.e. Coordinator and the Simulator that corresponds to a coupled model and an atomic model respectively. These components need to deploy at remote nodes so that distributed execution can take place.

In the current SOA framework, the Server can only acts as a provider of service and the Client only acts as a consumer of service. Contrary to this functionality, the DEVS simulation components mentioned above can be placed anywhere on the network. It is unavoidable that the same Server can act as a provider and a consumer while executing DEVS simulation protocol. Consequently, the SOA that executes the DEVS simulation protocol is constructed such that the servers that provide DEVS Service can play the role of both the Coordinator and the

---

kind of simulator -as long as the service updates the simulation cycle according to the simulator engine selected. The service is independent in the sense of transition functions.

Simulator. As shown in Figure 8, Step 2 provides a list of resources (servers) available on the Internet that provides DEVS simulation services. Once the list of servers is available to the User, he assigns the role of Coordinator to one of the servers and rests of them become Simulators. More details on this assignment is provided in Section 5.3.

During the execution of DEVS simulation protocol, each of the Simulators makes calls to other Simulator. Such calls are executed using the SOA framework. These Simulators also coordinate with the Coordinator using the same transport mechanism. As a result, the same Simulator is invoking services from other Simulators while providing services to other Simulators or Coordinator. This has resulted in an architecture that is symmetrical by default i.e. it acts as both a service provider and a service consumer. The temporal role of a remote node is guided by the DEVS simulation protocol.

The DEVS simulation layer services are defined in a separate WSDL that implements this symmetrical execution. Further, in addition to the roles of Simulator consumer and provider, the architecture allows the remote node to act as either Coordinator or Simulator. This assignment is made at Step 3 in Figure 8, and is elaborated in Section 5.3.

The next few sections give detailed account of the symmetrical server and client designs that implements symmetrical services architecture.

## 5.2 Server Design

### 5.2.1 Conceptual Design

#### 5.2.1.1 Abstraction of a Coupled Model with an Atomic Model with DEVS State Machine

One of the significant development steps we undertook in this effort is the masking of coupled model as an atomic model. Due to closure under coupling of the DEVS formalism, we have an abstraction mechanism by which a coupled model can be executed like an atomic model. In contrast to the DEVS hierarchical modeling, where a coupled model is merely a container and has corresponding coupled-simulators (Figure 9), now it is considered an atomic model with lowest level atomic simulator (Figure 10). This has been accomplished by implementing an adapter as shown in Figure 10. The adapter *Digraph2Atomic* takes each coupled component of the model and uses it as an atomic model.

The number of simulators created depends on the number of components of the model at the top-level and the number of servers selected by the user. If the model contains 10 top-level components (including the contained digraphs) and the user select 5 servers, then 2 simulators are created in each server. After the whole simulation process, each simulation service sends a report back to the user containing information related to IP addresses and simulator assignment.

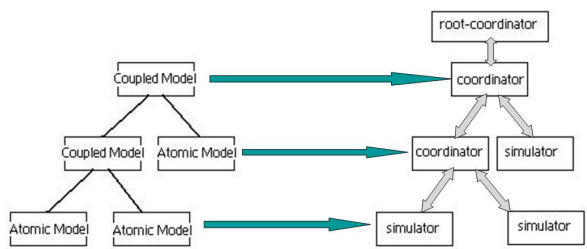

**Figure 9:** Hierarchical simulator assignment for a hierarchical model

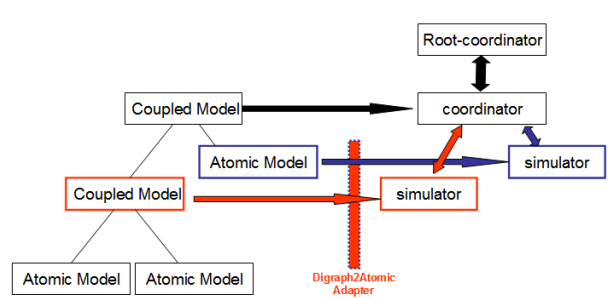

**Figure 10:** Hierarchical simulator assignment with Digraph2Atomic adapter

#### 5.2.1.2 Message Serialization
The issue of message passing and models upload is done through serialization and SOA technologies. Figure 7 illustrates the message serialization process. When a component makes an external transition or executes the output function, the message received or emitted is serialized and then sent to the coordinator through the simulation service. The coordinator stores the location of each simulation service, so he is able to request all the messages after each iteration.

All the communication between the coordinator and simulation services is done through SOA protocol. The serialization is done through Java serialization utilities. In a newly developed real-time version, each simulator knows each simulation service at its end (from coupling information). So the communication can be solved by passing messages from simulation services to simulation services directly, without using the coordinator.

#### 5.2.1.3 Centralized Simulation

The centralized simulation is done through a central coordinator which is located at the main server. The coordinator creates $n$ simulation services over Internet. Each simulation service creates $m$ simulators in order to simulation components of the model. Figure 11 shows the process. Once the simulation starts, the coordinator executes the output function of the simulation services (in Figure 11: point 0 and 1). After that, the output is propagated and internal transitions occur. Propagating an output means that once the coordinator takes the serialized output from the simulation services (2 and 3), it is sent to other simulation services by means of coupling information (4 and 5). This information is known by the coordinator and no others as all messages must flow through the coordinator.

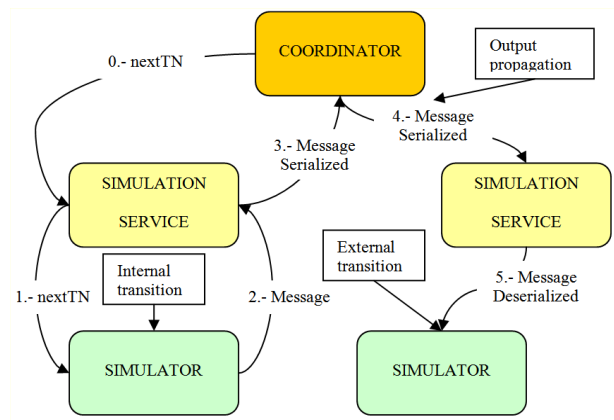 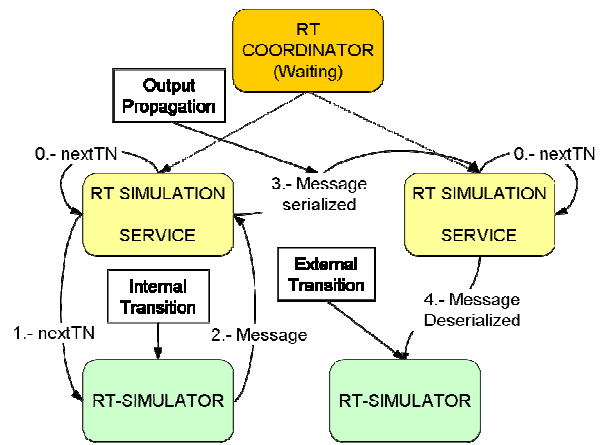

**Figure 11:** Centralized communication among services       **Figure 12:** Real-time communication among services

As it appears, the coordinator participates in all message-passing and is the bottleneck. We designed distributed DEVS SOA protocol where the coupling information is downloaded to each of the models and coordinator is relieved of message-passing. It is described as follows.

#### 5.2.1.4 Real-time Simulation

Real-time (RT) DEVS simulation is defined as the execution of DEVS simulation protocol in wall-clock time rather than logical time. For the real-time (RT) simulation we have incorporated one additional service to our SOA framework: the RT simulation service. This service extends the previous simulation service by means of two functions:
- Modify external output function
- Start simulation

The design is similar in many aspects, but instead of a central coordinator, all the simulation is observed by an RT coordinator without any intervention. Furthermore, the RT simulation service creates RT simulators. Each RT simulation service knows the coupling information, so the message passing is made directly from simulation service to simulation service at the other end. The RT coordinator is located at the main server. This coordinator creates $n$ RT simulation services over the Internet. Each simulation service creates $m$ RT simulators in order to simulate the components of the model. After that the coupling information is broken down (on a per-model basis) and sent to the corresponding RT simulation service. Figure 12 illustrates the process. Once the simulation starts, the coordinator executes the *simulate* service and nothing else. The simulate service waits for internal or external transitions using real time (0). If an internal transition happens (1), the output is generated and propagated using the coupling information serializing and de-serializing messages (2,3 and 4).

### 5.2.2 Package Design

The global design of the whole architecture at server's end is as follows, as shown in Figure . The **modeling** package constitutes the DEVS modeling library. Once a DEVS model is received by the servers, it is rebuilt using an adapter pattern. Presently, only DEVSJAVA models are allowed. But, since this framework follows an adapter pattern, other Java-based models will be allowed in future. Figure 14 depicts the classes contained in this package, such as *Digraph2Atomic*, *RTCoupling* for real-time simulation purposes and *Message* and *Atomic* classes. Both *Message* and *Atomic* classes are inherited from *Entity* which allows serialization and deserialization. *Atomic* encapsulates a DEVS atomic model and *Message* encapsulates a DEVS message or event.

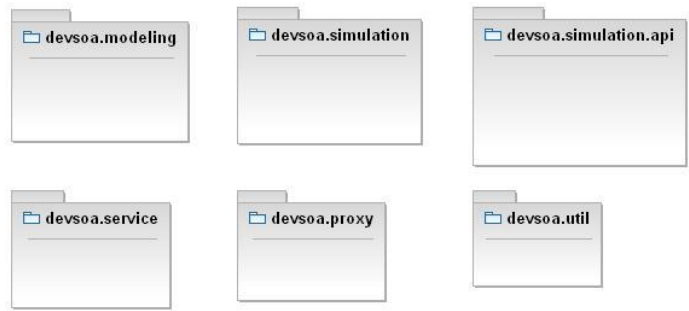

**Figure 13:** Server's package structure for DEVS SOA

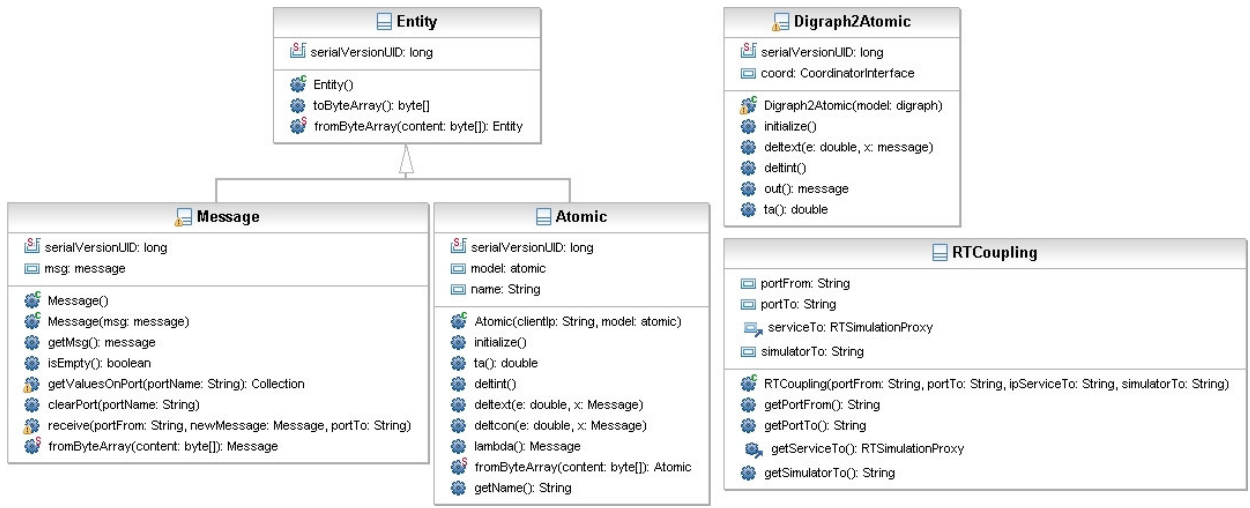

**Figure 14:** Modeling package for DEVS SOA

The **simulation.api** package contains the interface for our DEVS/SOA simulators. The **simulation** package contains simulators and coordinators, that is, *Simulator*, *Coordinator*, *RTSimulator* and *RTCoordinator* classes as shown in Figure . The *RT* prefix indicates that the class is designed for real-time simulation. The main difference with other simulators platforms starts here. In both centralized and real-time simulations, the Coordinator is executed at the first server selected by the user. This coordinator is called through a *MainService* class published as a Web service. The Coordinator receives the user IP, the name of the root coupled model, and a list of IPs. Such list of IPs is used to invoke simulation services in other remote servers. In this way, the components of the model are shared among N servers, where N is the length of that list. The Coordinator also stores the user IP, the DEVSJAVA model and a list of simulation services activated. In the case of centralized simulation, this list is used to propagate and to receive messages through the coupling protocol stored in the root coupled model. In addition, the Coordinator stores the last event time and the next event time. In the case of real-time simulations, instead of event times, the RTCoordinator only knows the time in which the simulation must be stopped.

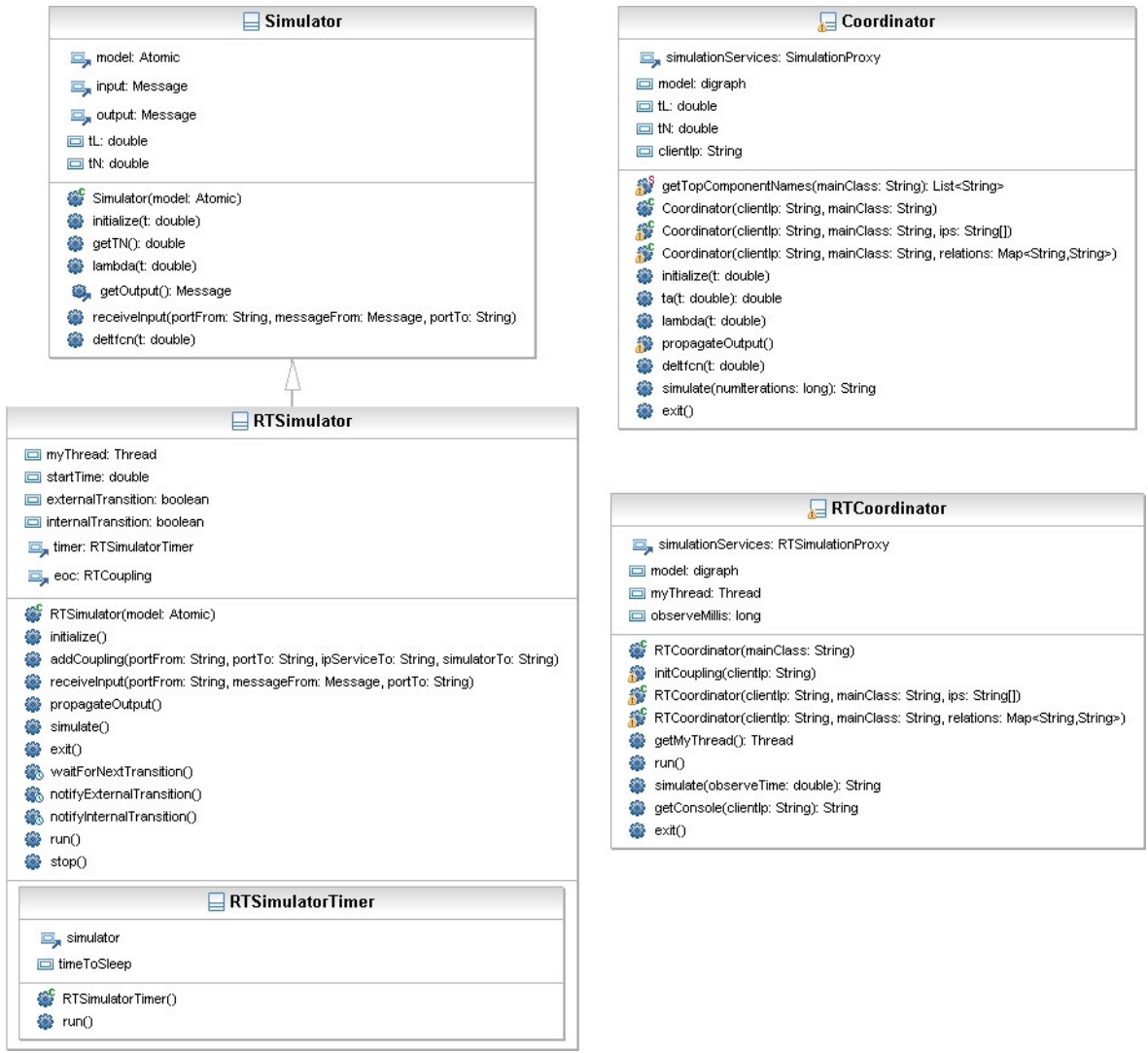

**Figure 15:** Simulation package in DEVS SOA

The **service** package contains the services offered. It contains *MainService*, *Simulation* and *RTSimulation* classes as shown in
Figure 16. *MainService* is designed to allow upload, compile and start the simulation process creating the coordinator. Simulation services are used to store the simulators used and to establish a communication between the DEVS simulators stored at this server and other coordinators, if any, hosted in other servers. One server could be executing more than one simulator. It depends on the number of components that the root coupled model contains and the number of servers selected by the user. This is the reason because there is not a unique relation between simulation service and simulator. The assignment of simulators corresponding to the models at the top-level is done by default through round-robin mechanism that takes care of model-simulator number mismatch. In certain applications, it is important that the user or a higher level program be able to direct any specific model to any particular IP server. For example, we are developing applications where DEVS models act as observers of co-hosted clients of other services. Clearly, ability to assign models to servers is critical in such an application.

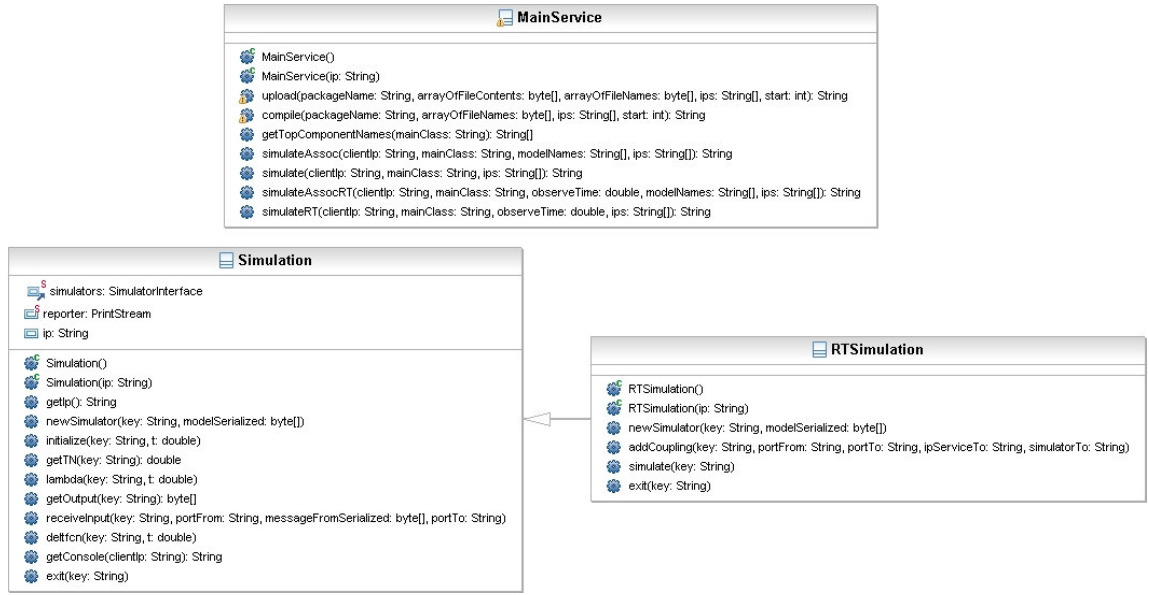

**Figure 16:** Service package in DEVS SOA

The **proxy** package (Figure 17) contains the proxies of the services. All these classes are automatically generated from the WSDL files that are generated from the service package using Apache Axis framework. The user only needs the *MainService* proxy. The server needs this service and other *Simulation* services. *MainService* acts like a coordinator for all the lower-level services through interfaces. It assigns and initializes the coordinator that starts other simulators, after distributing the simulators at respective IPs and initializing the simulator services. Once the simulators are active, the *MainService* waits for them to complete the execution to receive the logs and simulation outputs.

### 5.2.3 Symmetrical Service Design

The simulation engine is implemented in two different ways. The first is the centralized version with logical time execution and the other is a real-time version. The details below cater to the centralized version. The operations of real-time version are almost the same except that instead of just the coordinator controlling the simulation clock, each of the simulators maintains its own thread in real-time and exchange messages independently without the intervention from coordinator.

As described earlier, this framework is a layered framework containing two layers:
1. User Layer
2. Simulation Layer

The User layer is called as *MainService* layer and it interfaces with the Simulation layer underneath. The user can freely consider both the centralized and distributed version of the simulation algorithm. This facility is provided at the second layer of services described in later sections. However, the centralized mode performs much slower than the real-time distributed simulation due to obvious reasons of coordinator loading.

In developing DEVS/SOA client, we considered real-time simulation as the default option. Detailed performance analysis of both of these implementations is in process and will be reported in our forthcoming publication.

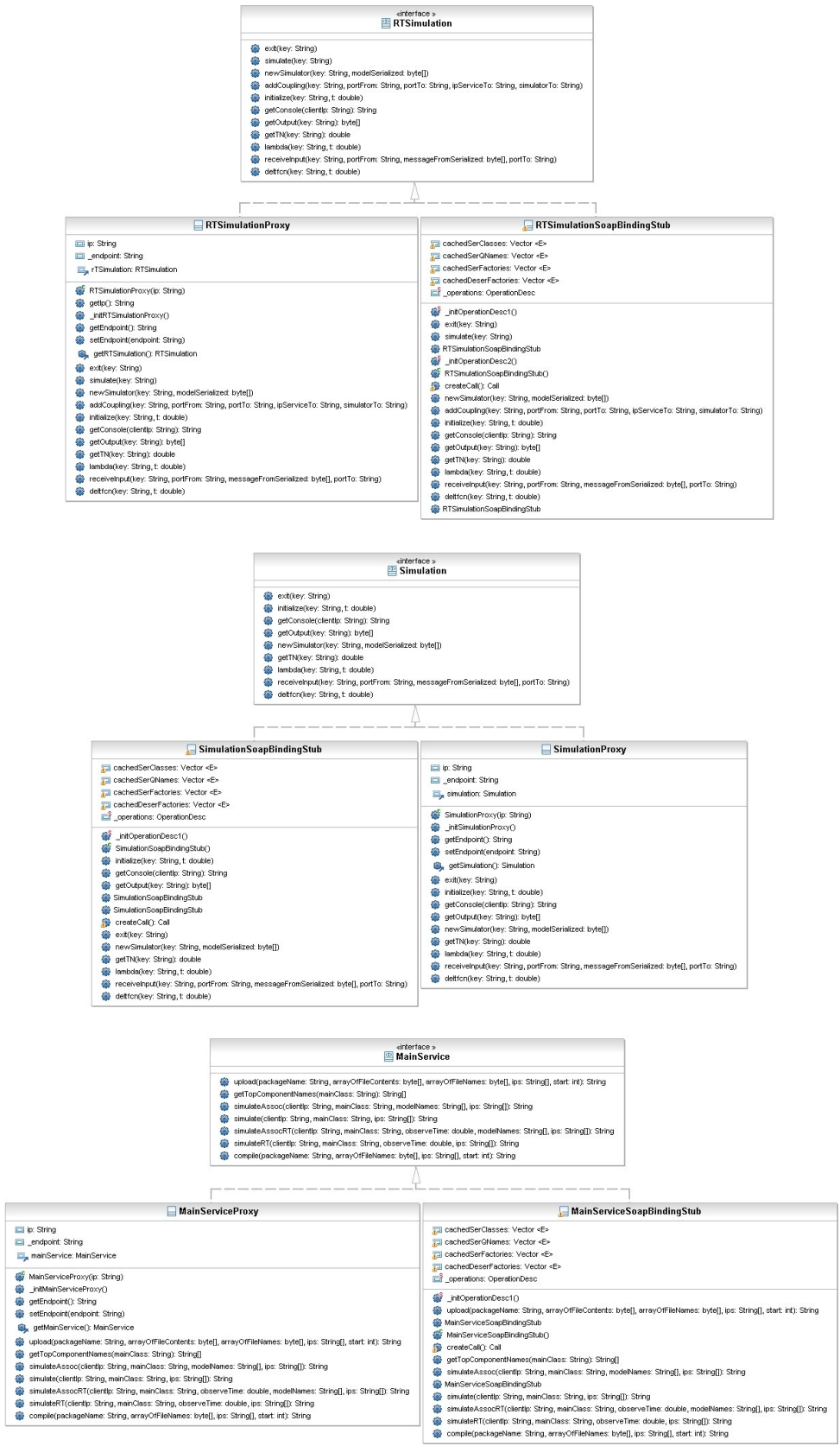

**Figure 17:** Proxy package in DEVS SOA

#### 5.2.3.1 *MainService* composition:

The *MainService* layer provides the set of services that are available to the user (as a client). The **MainService.wsdl** is provided in the appendix that the user can use to implement its own client. For better usability, we have implemented the Client as well and it is described in the following sections. The *MainService* layer provides the following services:

- *upload*: It is used to upload the model to the different servers. This service enables the user to take their DEVS models and upload the code physically from their machines to the designated DEVS/SOA server farm. This service receives (1) the package name, which is the folder where the model is saved at the server side, (2) the content of the java files, which is in fact the DEVS model implementation, (3) the name of the java files, and (4) the list of IP addresses where the model is being uploaded. Once the model is uploaded to the first server of the list, the server application executes this service in the next server of the list.
- *compile*: This service is used to compile the uploaded model files at the Servers and make them ready to execute the simulation. It receives (1) the package name, which is the folder where the model was previously uploaded, (2) the file names of the DEVS model implementation, and (3) the list of IP addresses that are selected by the user and where the model is simulated. In the client application we have developed, the first argument is dynamically generated at the client's end and is important because if the model is uploaded with the same package name repeatedly, the server class loader does not instantiates the last one compiled. To overcome this issue, the model files do not contain any package declaration and a package name is assigned at run-time compilation. Once the model is compiled at the first server's end, the server application executes this service in the next server of the list of IP addresses.
- *getTopComponentNames*: This service is used to obtain the name of the top-level DEVS model. It receives the name of the root coordinator and returns the array of names. This service may be used to associate an IP server address with each of the top-level DEVS components.
- *simulate, simulateAssoc, simulateRT and simulateAssocRT*: The simulation services create a Coordinator which runs its *simulate* function. The *RT* suffix indicates that a real-time simulation service is required by the user. The *Assoc* suffix indicates that the user is passing relations (IP address, model's name), that is, in which server the corresponding model is executed. The non-*Assoc* functions apply a round-robin algorithm. The main difference among these functions is the coordinator created. A *Coordinator* in the case of centralized simulation and an *RTCoordinator* in the case of real-time simulations (see Figure 11,12). In all cases, such services receive: (1) the IP address of the client running the service, (2) the name of the root-coordinator in the DEVS model, and (3) the relation between model names and IP addresses (if it is provided by the user). In the case of real-time simulation, the service also receives the time to observe the simulation. Finally, the service returns the simulation results.

#### 5.2.3.2 *Simulation* service composition:

**This is the bottom layer of the two-layer** architecture and its functionalities are used by the *MainSevice* layer. Its operations are transparent to the user. Once the user demands a simulation via the *MainService* class, the coordinator (at the coordinator server or main server) requires as many simulation services as IP addresses provided by the user. After that, the DEVS model is partitioned and the coordinator sends every part to its corresponding service. Then the simulation starts, each simulation service creates a DEVS simulator for its models and executes the corresponding output and transition functions (see Figure 11).

It is possible for one simulation service to store more than one simulator for different component of the same DEVS model, or to store more than one simulator for different components of different DEVS models. This issue is solved as follows. After the main coordinator obtains a simulation service at a certain IP address, a new simulator is created there, identified by the component name plus the IP address of the user's machine and containing the DEVS component itself. For example, if the coordinator must send a DEVS component named *Processor* to a server located at 192.168.1.5 and coming from a user located at 192.168.1.2, then a simulation service is required from 192.168.1.5 and a new simulator is created there, identified by *Processor@192.168.1.2* and containing the model named *Processor*.

Another issue is how to store the simulators created, because web services do not have memory. To this end, we are using the server's memory by means of static variables or attributes. Hence, the simulation services include a static table, which associates *simulator names* with *simulator instances*. Figure 15 shows this attribute in the Simulation service class, called *simulators*.

There is other information stored by the Simulation services in the server memory, such as the IP address where the services reside and a reporter, which logs all the information while the simulation is running.

The services provided by the Simulation service are enumerated below:
- *newSimulator:* This service receives a DEVS component and a identifier. It creates a new DEVS simulator identified by the name described above and containing the DEVS component received.
- *initialize:* This service receives the name of the simulator required and the current time. It takes the corresponding simulator from its table (using the name received) and initializes it.
- *receiveInput:* This service receives four arguments: (1) the name of the simulator required, (2) the name of the port where the message is coming from, (3) the message and (4) the name of the port where the message is going to. The simulation service takes the simulator from its table and executes the same function called *receiveInput*, which stores the message received at the input of the model.
- *lambda:* It receives the name of the simulator required and the current time. This service takes the simulator required and executes the output function (also called *lambda)* of the DEVS model
- *deltfnc:* This service receives the name of the simulator required and the current simulation time. The service takes the simulator and executes an internal or external or confluent transition function. The abstracted deltfn is provides in Figure 18. This allows both the classical DEVS and P-DEVS models work seamlessly with DEVS/SOA simulation framework.
- *getOutput:* This service receives the name of the simulator required and returns the output stored in its DEVS model.
- *getTN:* It receives the name of the simulator for which the time of the next event is returned.
- *exit:* It receives the name of the simulator to be removed from the table.
- *getConsole:* This service receives the IP address of the user's machine, and return the content of the log file related to this address.
- *getIp:* It returns the IP address of the simulation service.

```
function deltfcn(double t) {
        Message x = input;
        if(x==null) {
                System.out.println(
                "ERROR RECEIVED NULL INPUT " + model.toString());
                return;
        }
        if (x.isEmpty() && t!=tN) {
                return;
        }
        else if((!x.isEmpty()) && t==tN) {
                double e = t – tL;
                model.deltcon(e,x);
        }
        else if(t==tN) {
                model.deltint();
        }
        else if(!x.isEmpty()) {
                double e = t – tL;
                model.deltext(e,x);
        }
        tL = t;
        tN = tL + model.ta();
        input = new Message();
}
```

Figure 18: Abstract deltfun in Simulation service

Having described the services available in the DEVS/SOA architecture, following is the design of DEVS/SOA coordinator and simulator that utilize these DEVS services. The coordinator and the simulator are implemented in the *devsoa.simulation* package. This simulator is called as DEVSV/SOA simulator and it acts as an adapter for any DEVS simulation engine that executes the DEVS simulation protocol. Currently, it adapts to the DEVSJAVA Version 3.0 as available from ACIMS.

**DEVSV/SOA Coordinator:**

Equivalent to the Simulation service storing the simulators in a static way, the coordinator also stores the simulators of the DEVS model in a static hash table, using the same nomenclature as was stated above (DEVS component name plus client IP address identifying the simulator). Therefore, such table contains pairs {simulator name, simulator service}, associating each simulator created with the simulation service where it resides. The task of the coordinator is to execute a typical DEVS loop over the distributed simulators. Figure 19 shows the algorithm executed by the *simulate* function. In such Table, *iterations* is the number of cycles of the simulation, *t* is the current time, *tL* is the last time event, *tN* is the next time event, *simulationServices* is the table of simulation services created by the coordinator and where the simulators are located. Then, for a number of cycles, the output function is called through each of the simulation services. It should be noted that the first argument of *lambda* function is a key, which is the simulator identifier, since different simulators could be located at the same simulation service, this key must be provided. After the output function is executed, the outputs of the components are ready to be propagated. To this end, the *propagateOutput* function is called, which propagates the messages generated from the outports to its corresponding inports. Next, the transition function is applied and finally the time is updated.

```
function simulate(long iterations)
  t = tN;
  for (i=0; i<iterations; i++)
    for each ({key,simService} in simulationServices)
      simService.lambda(key, t);
      propagateOutput();
    for each ({key,simService} in simulationServices)
      simService.deltfcn(key, t);
    tL = t;
    tN = min(simulationServices.getTN());
    t = tN;
```

**Figure 19:** DEVS simulation

From the instant in which the coordinator is created, it stores at any moment the DEVS model (currently DEVSJAVA), the last timed event, the next time event and the IP address of the user's machine.

It should be noted that the Coordinator is not a service. It is a class, which is used by the *MainService* service. The functions implemented in the Coordinator are enumerated below:

- *getTopComponentNames:* This function receives the name of the DEVS root-coupled model and returns a list containing the top-component names of the DEVS model.
- *Constructor:* The constructor receives the client IP address, the name of the DEVS model, and the list of IP addresses where the model is going to be simulated. Hence, it creates as many simulators as top-level components, created by the simulation services located at the IP addresses given in the list.
- *initialize:* This function receives the initial time of simulation. It initializes the simulators.
- *propagateOutput:* As it was stated above, this function takes the output from the simulators and sends them to its corresponding inputs.
- *lamda:* It receives the current time, and executes the output function in each of the simulators stored.
- *deltfcn:* This function receives the current time and executes the internal or external transition functions in the simulators stored.
- *ta:* It is the time advance function and receives the current time. It takes the minimum next time event from the simulators stored.
- *exit:* This function calls the exit function of all the simulation services stored and clean the table of simulators.
- *simulate:* This function receives the number of cycles of the simulation, and executes the simulation as was described before (Figure 19).

## 5.3 Client Application

This Section provides the client application to execute DEVS model over an SOA framework using Simulation as a Service. From many-sided modes of DEVS model generation (Figure 7), the next step is the simulation of these models. The DEVSV/SOA client takes the DEVS models package and through the dedicated servers hosting simulation services, it performs the following operations:
1. Upload the models to specific IP locations
2. Run-time compile at respective sites
3. Simulate the coupled-model
4. Receive the simulation output at client's end

The DEVSV/SOA client as shown in Figure 20 operates in the following sequential manner:
1. The user selects the DEVS package folder at his machine
2. The top-level coupled model is selected as shown in Figure 21.
3. Various available servers are selected (Figure 21). Any number of available servers can be selected (one at least).
4. Clicking the button labelled "Assign Servers to Model Components" the user selects where is going to simulate each of the coupled models, including the top-level one, i.e., the main server where the coordinator will be created (Figure 21)
5. The user then uploads the model by clicking the Upload button. The models are partitioned and distributed among the servers chosen in the previous point
6. The user then compiles the models at the server's end by clicking the Compile button

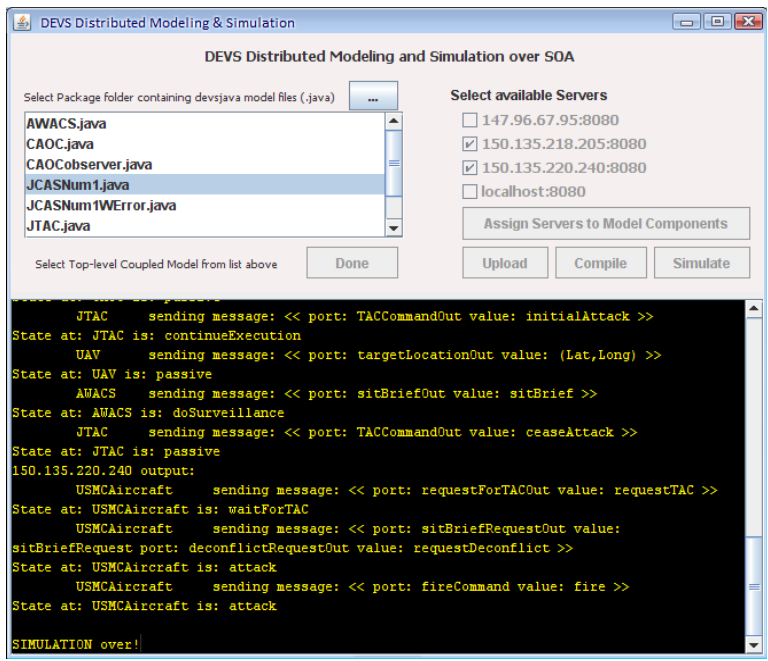

**Figure 20:** GUI snapshot of DEVSV/SOA client hosting distributed simulation

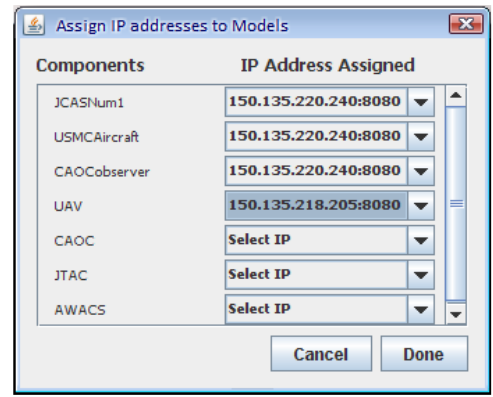

**Figure 21:** Server Assignment to Models

## 6. Cross-Platform Execution over DEVS/SOA

### 6.1 Introduction

In terms of net-ready capability testing, what is required is the communication of live web services with those of test-models designed specifically for them. The approach we are working on has the following steps:
1. Specify the scenario
2. Develop the DEVS model
3. Develop the test-model from DEVS models
4. Run the model and test-model over SOA
5. Execute as a real-time simulation
6. Replace the model with actual web-service as intended in scenario.
7. Execute the test-models with real-world web services
8. Compare the results of steps 5 and 7.

Of course, many issues of policy management and security considerations must be taken care of when test-models are communicating with live Web-Services. However, considering the fact that for any defense related mission-thread reliability testing the test-models would have the necessary security provisions, the 8-step process listed above can be executed. This work would also involve generation of DEVS models from WSDLs specifications. A small portion of BPMN-to-DEVS transformation is described in [31].

One other section that requires some description is the multi-platform simulation capability as provided by DEVSV/SOA framework. It consists of realizing distributed simulation among different DEVS platforms or simulator engines such as DEVSJAVA, DEVS-C++, etc. In order to accomplish that, the simulation services will be developed that are focused on specific platforms, however, managed by a coordinator. In this manner, the whole model will be naturally partitioned according to their respective implementation platform and executing the native simulation service. This kind of interoperability where multi-platform simulations can be executed with our DEVSML integration facilities. DEVSML will be used to describe the whole hybrid model. At this level, the problem consists of message passing, which has been solved in this work by means of an adapter pattern in the design of the "message" class (used in Figure 11 and 12). Figure 22 shows a first approximation. The platform specific simulator generates messages or events, but the simulation services will transform these platform-specific-messages (PSMsg) to our current platform-independent-message (PIMsg) architecture developed in DEVS/SOA.

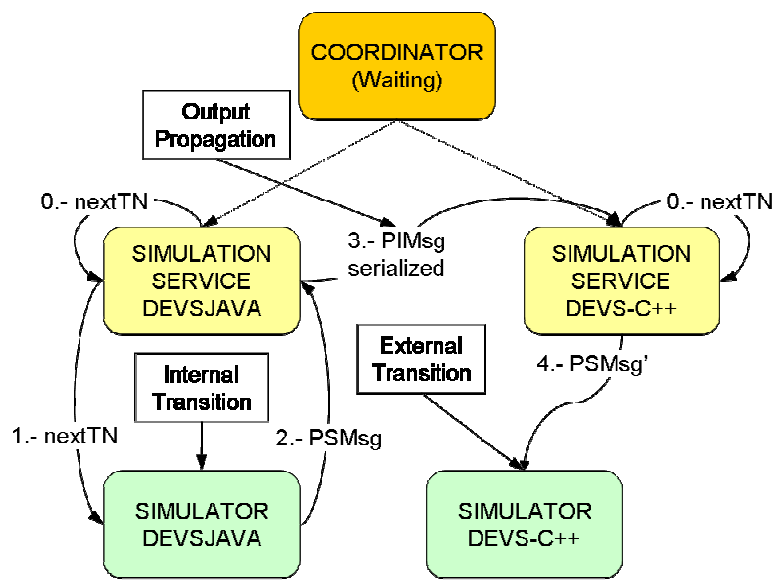

**Figure 22:** Cross-platform execution. First approximation

Hence, we see that the described DEVS/SOA framework can be extended towards net-ready capability testing. The DEVS/SOA framework also needs to be extended towards multi-platform simulation capabilities that allow test-models be written in any DEVS implementation (e.g. Java and C++) to interact with other as services.

However, a major drawback of our current architecture is that the user must send the whole DEVS model implemented under all the platforms to use, which is not a good solution. Next, we propose a modification on the Coordinator creation process that in some manner, allows to the user to store each part of the model written in its corresponding platform.

## 6.2 Multi-platform DEVS/SOA architecture

Figure 23 depicts an example of a multi-platform DEVS model. Each atomic or coupled component may be implemented using different simulation engines, called *platforms*. In Figure 23, SUBMODEL A is implemented using DEVSJAVA [9], SUBMODEL B by means of aDEVS (C++) [38], and SUBMODEL C using xDEVS (Java) [4].

Let us suppose that the whole model is implemented using DEVSJAVA. In our current DEVS/SOA architecture, the application sends the whole model (root-coupled model included) to the servers by means of the *upload* service, where all the files get compiled and finally, it executes the model sending serialized messages among simulation services. This situation is not valid for the multi-platform model depicted in
Figure , since the scenario cannot be compiled as a whole.

In our proposed approach, we define the root coordinator by means of a *Platform Independent Model (PIM)*, for example, DEVSML. We may use the structure description of DEVSML to compose the root coupled model, and send it to the main server, which will distribute the sub-models among its corresponding servers.

Figure shows how a multi-platform DEVS model may be executed using our proposed architecture. We define the root-coupled model using DEVSML (top of the Figure 24). The coupled model is treated as an atomic model due to the inherent architecture of DEVS/SOA *digraph2Atomic* adapter. Consequently, it is immaterial if the sub-model is atomic or coupled (Section 5.2.1.1).

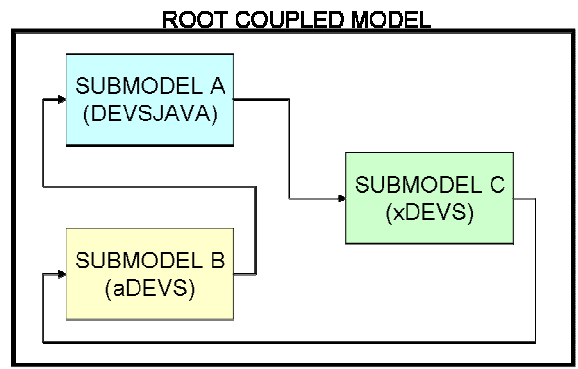

**Figure 23:** Multi-platform DEVS model

The DEVSML document in the Figure 23 states that the main server is located at 192.168.1.3. This server receives the DEVSML document and all the source code, distributes sub-models to respective servers and creates the coordinator. For example, the main server sends *SubModelA.java* to the server located at 192.168.1.7, where the DEVS/SOA java implemented server compiles it. The same happens with the corresponding *SubModelB.cpp* and *SubModelC.java*. After compiling all sub-models, the main server creates one simulation service for each sub-model. Figure 24 (right side) shows how coordinator, simulation services, and simulators are created. The main server creates a DEVSJAVA-based simulation service located at 192.168.1.7, which also creates a DEVSJAVA-based simulator to store *SubModelA*. The same occurs with sub-models B and C, but at IP addresses 192.168.1.5 and 192.168.1.9 respectively.

The rest of the behavior of the application is the same that in our current architecture. Messages are passed by means of an adapter pattern, which as Figure depicts, may be translated into different platforms.

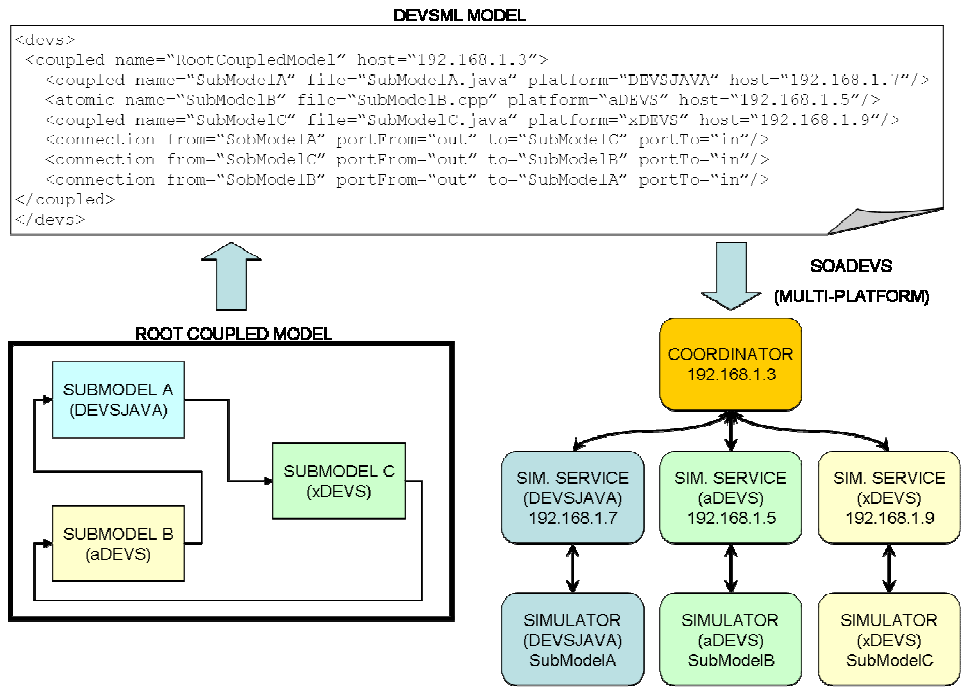

**Figure 24:** Multi-platform DEVSV/SOA proposed architecture

## 7. Applications

This section contains two sub-sections. Section 7.1 deals with an example in lab-setting through which various concepts laid out in earlier sections are demonstrated. Section 7.2 brings about a real-world application that could utilize the capabilities provided by DEVS/SOA framework.

### 7.1 Joint Close Air Support Example

The JCAS system requirements come in many formats and it served as a base example to test many of the DUNIP earlier processes for requirements-to-DEVS transformation. It was specified using the state-based approach, BPEL-based approach and restricted natural language approach [31]. This case study describes all three of the approaches leading to an executable DEVS model with identical simulation results. Finally, the executable model is executed over a net-centric platform using DEVSML and DEVS/SOA architecture.

The Joint Close Air Support Model is expressed in plain English as shown in Figure 25 below. It is a small example involving components exchanging messages towards a common objective. The requirements are then translated to various DEVS generating modes. We shall see the execution of JCAS for each of the approaches. The components of JCAS model are:
1. JTAC
2. UAV
3. CAOC
4. USMC Aircraft
5. AWACS

The scenario is as follows:

| JCAS JMT Operational Scenario #1 |
| --- |
| A. Special Operations Force (SOF) (AFSOC and NSW) JTAC working with Operational Detachment-Alpha (ODA) is tasked to request Immediate CAS on a stationary mechanized target in mountainous terrain.  A Predator unmanned aerial vehicle (UAV) is on station for support. |
| B. SOF JTAC contacts AWACS with request.  AWACS passes the request to Special Operations Liaison Element (SOLE) in the Combine Air Operations Center (CAOC). |
| C. Joint Special Operations Task Force (JSOFT) approves the request and CAOC assigns a section of USMC F/A-18Ds, F-15Es, and a single B-1B.  Ordnance consists of 20mm, Joint Direct Attack Munitions (JDAMs), and Laser Guided Bombs (LGBs). |
| D. Aircraft get situational brief from AWACS aircraft while in route, then switch to SOF JTAC for Terminal Attack Control and deconfliction from orbiting UAV.  A 9-Line brief will be given to each section/single aircraft.  JTAC will continue to execute CAS missions until all weapons are expended. |

**Figure 25:** JCAS Operational Scenario

We approached the scenario using a BPMN diagram. The scenario in Figure 25 is expressed as a BPMN diagram shown in Figure 26 below. The BPMN diagram was created manually using the tool Borland Eclipse Together 2006. The Eclipse Together tool generated the corresponding .bpel and .wsdl files for the JCAS scenario. In total 10 files were generated (5 .bpel and 5 .wsdl files). The generated files are shown in Figure 27.

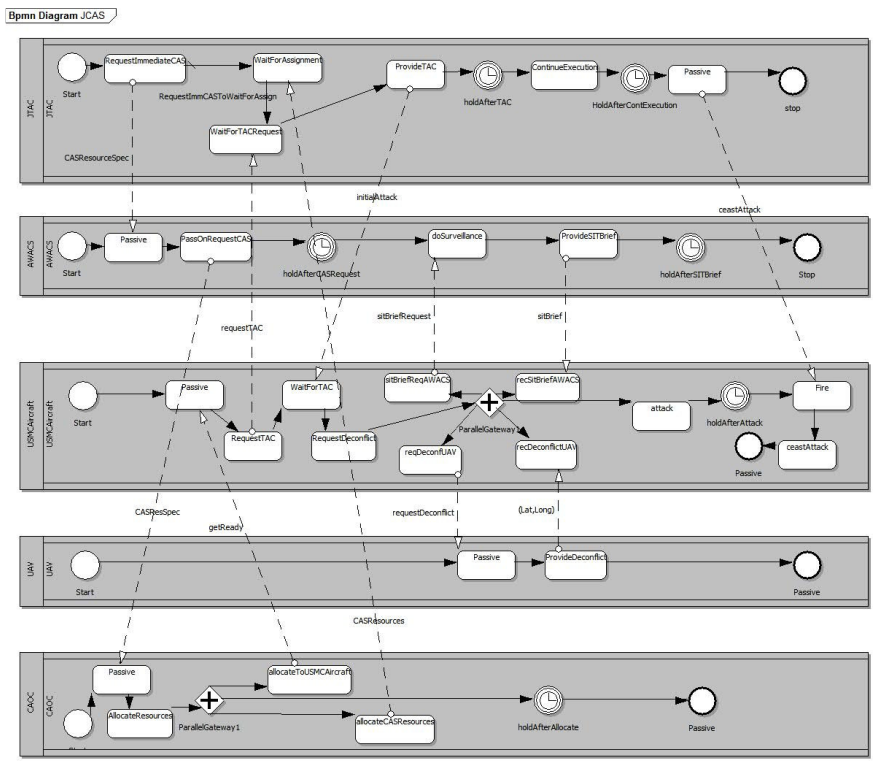

**Figure 26:** JCAS BPMN scenario description

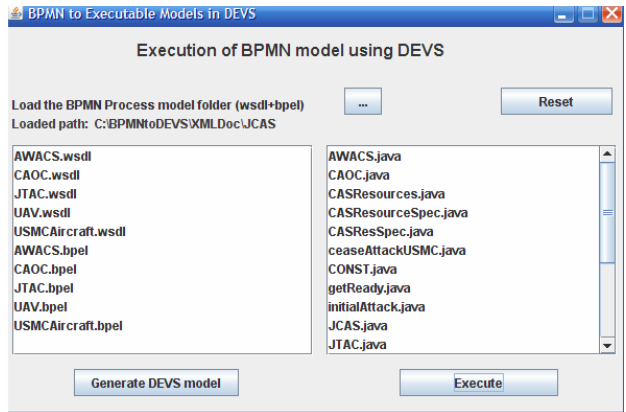

**Figure 27:** Snapshot of a BPMN-to-DEVS Transformation tool

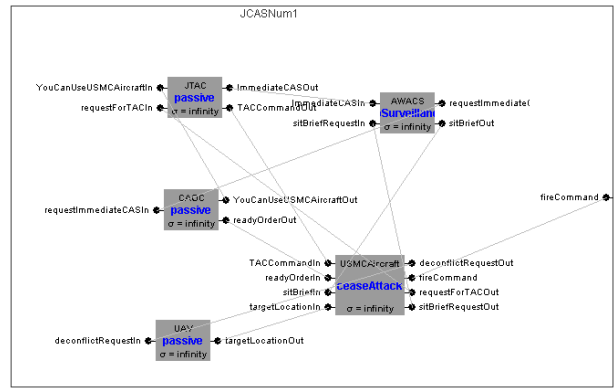

**Figure 28:** Coupled scenario for JCAS model

We took these generated files to our BPEL-to-DEVS transformation tool [31] and generated the DEVS model out of these files. The transformation process generated the following .java files (which include additional files as well) shown in Figure 27 above.

1. **JCAS.java**
2. **JTAC.java**
3. **AWACS.java**
4. **CAOC.java**
5. **UAV.java**
6. **USMCAircraft.java**
7. CASResources.java
8. CASResourceSpec.java
9. CASResSpec.java
10. ceaseAttackUSMC.java
11. CONST.java
12. getReady.java
13. initialAttack.java
14. latLong.java
15. requestDeconflict.java
16. requestTAC.java
17. sitBrief.java
18. sitBriefRequest.java
19. TimerMessage.java

The additional files correspond to various messages that were exchanged in the scenario. The files in the bold (above) are the main component files that contain the DEVS state machine.

Finally, using the BPMN-to-DEVS tool, the package was compiled run-time and simulation was executed. The Execute button brings up the DEVSJAVA Simulation Viewer (Figure 28) which executes the simulation.

**Net-centric Execution of JCAS**

Execution of JCAS DEVS models on net-centric SOA platform was done using the DEVS/SOA tool. The client application as described in Section 5.3 was used to execute the operation. Two servers were selected to demonstrate the concept (as shown in Figure 29). Both the servers are located at ACIMS lab, University of Arizona. However other server at Spain, University Computense de Madrid were also used in various testing sessions. Also shown in Figure 29 (in the console window) is the process of files being uploaded, compiled and the simulation-in-progress.

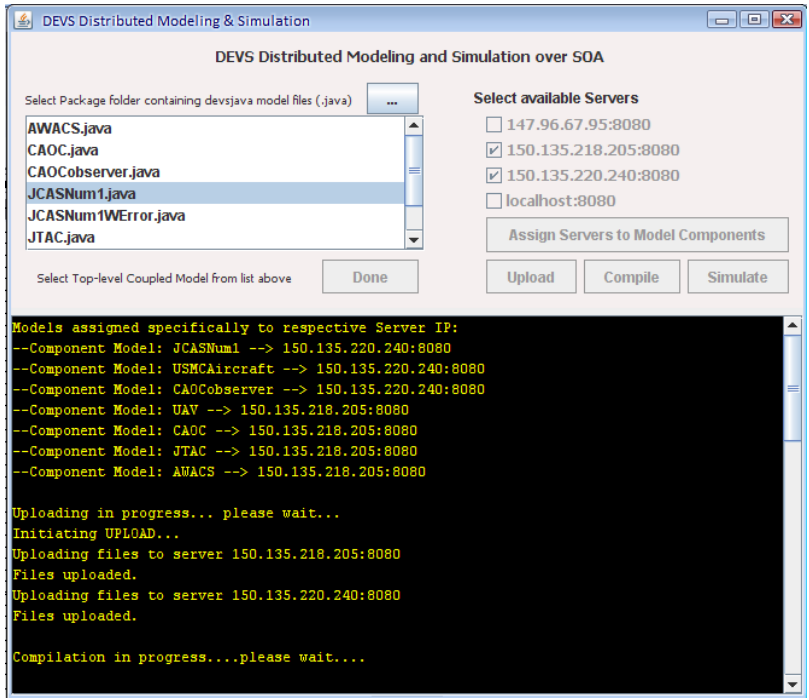

**Figure 29:** DEVSV/SOA client running the JCAS model using Simulation services on two hosts

Finally, when the simulation is over, the console displays the following output. The simulation logs from both of the servers are categorically displayed. Figure 30 below shows the complete console log for all the operations done using DEVS/SOA client.

```
Models assigned specifically to respective Server IP:
--Component Model: JCASNum1 --> 150.135.220.240:8080
--Component Model: USMCAircraft --> 150.135.220.240:8080
--Component Model: CAOCobserver --> 150.135.220.240:8080
--Component Model: UAV --> 150.135.218.205:8080
--Component Model: CAOC --> 150.135.218.205:8080
--Component Model: JTAC --> 150.135.218.205:8080
--Component Model: AWACS --> 150.135.218.205:8080

Uploading in progress... please wait...
Initiating UPLOAD...
Uploading files to server 150.135.218.205:8080
Files uploaded.
Uploading files to server 150.135.220.240:8080
Files uploaded.

Compilation in progress....please wait....

Starting compilation at remote servers.....
Compiling project at 150.135.218.205:8080...
Project compiled.
Compiling project at 150.135.220.240:8080...
Project compiled.

Waiting to start SIMULATION....

Simulation in Progress....please wait...
Running simulation ...
11 iterations.
Simulators output:

150.135.218.205 output:
        JTAC    sending message: << port: ImmediateCASOut value: CASResourcesSpec >>
```

```
State at: JTAC is: waitForAssignment
        AWACS    sending message: << port: requestImmediateCASOut value: CASResourcesSpec
>>
State at: AWACS is: doSurveillance
        CAOC     sending message: << port: readyOrderOut value: getReady port:
YouCanUseUSMCAircraftOut value: CASResources >>
State at: CAOC is: passive
        JTAC     sending message: << port: TACCommandOut value: initialAttack >>
State at: JTAC is: continueExecution
        UAV      sending message: << port: targetLocationOut value: (Lat,Long) >>
State at: UAV is: passive
        AWACS    sending message: << port: sitBriefOut value: sitBrief >>
State at: AWACS is: doSurveillance
        JTAC     sending message: << port: TACCommandOut value: ceaseAttack >>
State at: JTAC is: passive

150.135.220.240 output:
        USMCAircraft     sending message: << port: requestForTACOut value: requestTAC >>
State at: USMCAircraft is: waitForTAC
        USMCAircraft     sending message: << port: sitBriefRequestOut value:
sitBriefRequest port: deconflictRequestOut value: requestDeconflict >>
State at: USMCAircraft is: attack
        USMCAircraft     sending message: << port: fireCommand value: fire >>
State at: USMCAircraft is: attack

SIMULATION over!
```

**Figure 30:** Simulation output at client's application using DEVS/SOA client

## 7.2 Distributed Multi-level Test Federations

A DEVS distributed federation is a DEVS coupled model whose components reside on different network nodes and whose coupling is implemented through middleware connectivity characteristic of the environment, e.g., SOAP for GIG/SOA. The federation models are executed by DEVS simulator nodes that provide the time and data exchange coordination as specified in the DEVS abstract simulator protocol.

As discussed earlier, in the general concept of experimental frame (EF), the generator sends inputs to the SoS under test (SUT), the transducer collects SUT outputs and develops statistical summaries, and the acceptor monitors SUT observables making decisions about continuation or termination of the experiment [18]. Since the SoS is composed of system components, the EF is distributed among SoS components, as illustrated in Figure 31. Each component may be coupled to an EF consisting of some subset of generator, acceptor, and transducer components. As mentioned, in addition an observer couples the EF to the component using an interface provided by the integration infrastructure. We refer to the DEVS model that consists of the observer and EF as a *test agent*.

Net-centric Service Oriented Architecture (SOA) provides a currently relevant technologically feasible realization of the concept. As discussed earlier, the DEVS/SOA infrastructure enables DEVS models, and test agents in particular, to be deployed to the network nodes of interest. As illustrated in Figure 31, in this incarnation, the network inputs sent by EF generators are SOAP messages sent to other EFs as destinations; transducers record the arrival of messages and extract the data in their fields, while acceptors decide on whether the gathered data indicates continuation or termination is in order [31].

Since EFs are implemented as DEVS models, distributed EFs are implemented as DEVS models, or agents as we have called them, residing on network nodes. Such a federation, illustrated in Figure 32, consists of DEVS simulators executing on web servers on the nodes exchanging messages and obeying time relationships under the rules contained within their hosted DEVS models.

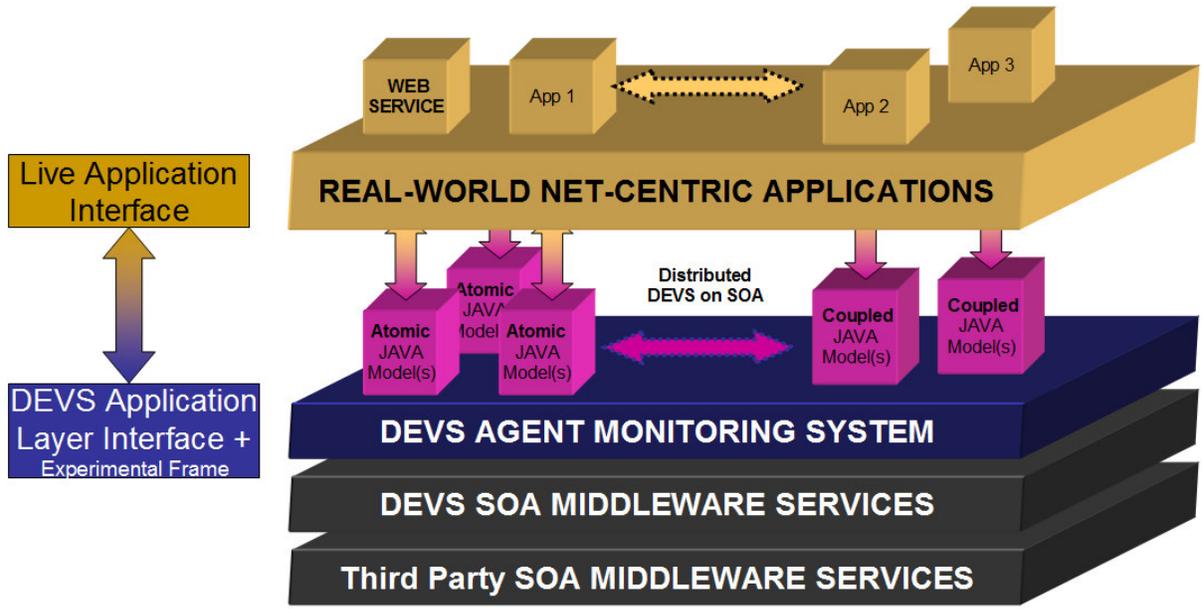

**Figure 31:** Deploying Experimental Frame Agents and Observers

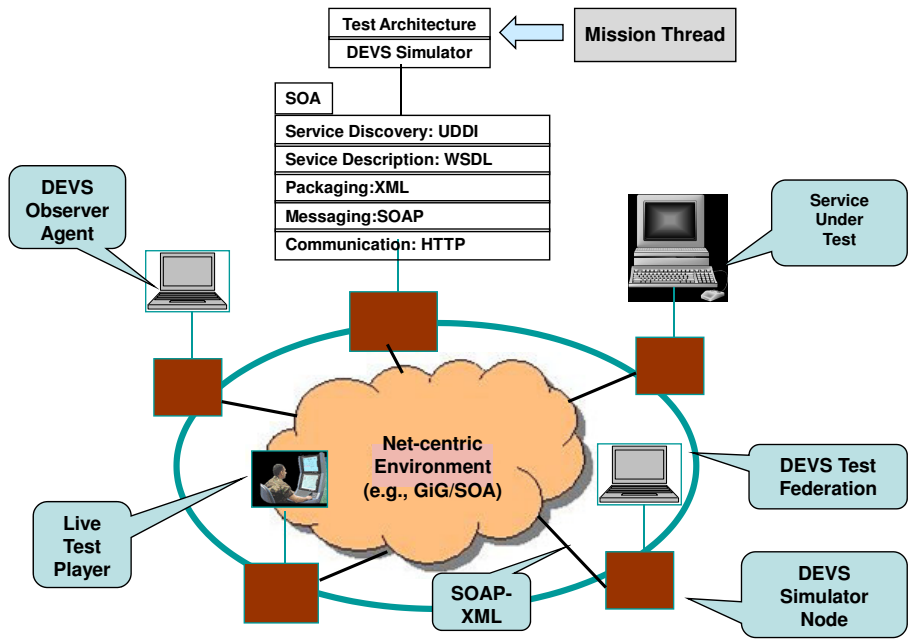

**Figure 32:** DEVS Test Federation in GIG/SOA Environment

The linguistic levels of interoperability [37] provide a basis for further structuring the test instrumentation system. In the following sections, we discuss the implementation of test federations that simultaneously operate at the syntactic, semantic, and pragmatic levels (Figure 33).

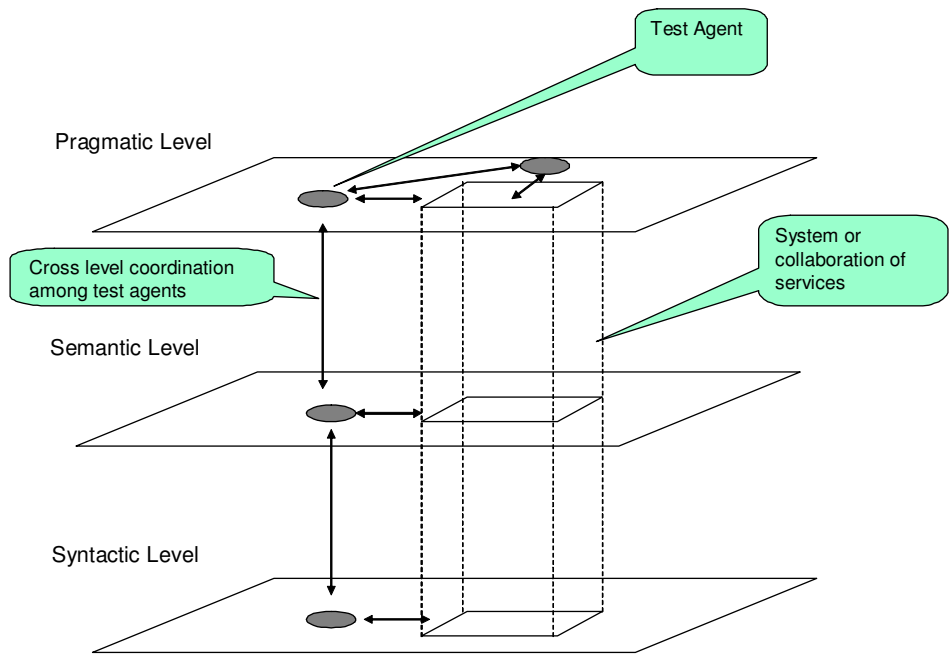

**Figure 33:** Simultaneous testing at multiple levels

### 7.2.1 Syntactic Level – Network Health Monitoring

From the syntactic perspective, testing involves assessing whether the infrastructure can support the speed and accuracy needed for higher level exchange of information carried by multimedia data types, individually and in combination. We now consider this as a requirement to continually assess whether the network is sufficiently "healthy" to support the ongoing collaboration. Figure 34 illustrates the architecture that is implied by the use of subordinate probes. Nodal generator agents activate probes to meet the health monitoring Quality of Service (QOS) thresholds determined from information supplied by the higher layer test agents, viz., the objectives of the higher layer tests.

Probes return statistics and alarm information to the transducers/acceptors at the DEVS health layer which in turn may recommend termination of the experiment at the test layer when QOS thresholds are violated. In an EF for real-time evaluation of network health, the SUT is the network infrastructure (OSI layers 1-5) that supports higher session and application layers. QOS measures are at the levels required for meaningful testing at the higher layers to gather transit time and other statistics, providing quality of service measurements.

For messages expressed in XML and carried by SOAP middleware such messages are directly generated by the DEVS generators and consumed by the DEVS transducers/acceptors. Such messages experience the network latencies and congestion conditions experienced by messages exchanged by the higher level web servers/clients. Under certain QOS conditions however, video streamed and other data typed packets may experience different conditions than the SOAP-borne messages. For these we need to execute lower layer monitoring under the control of the nodal EFs.

The collection of agent EFs has the objective of assessing the health of the network relative to the QOS that it is providing for the concurrent higher level tests. Thus such a distributed EF is informed by the nature of the concurrent test for which it monitoring network health. For example, if a higher level test involves exchanges of a limited subset of media data types (e.g., text and audio), then the lower layer distributed EF need only monitor the subset of types.

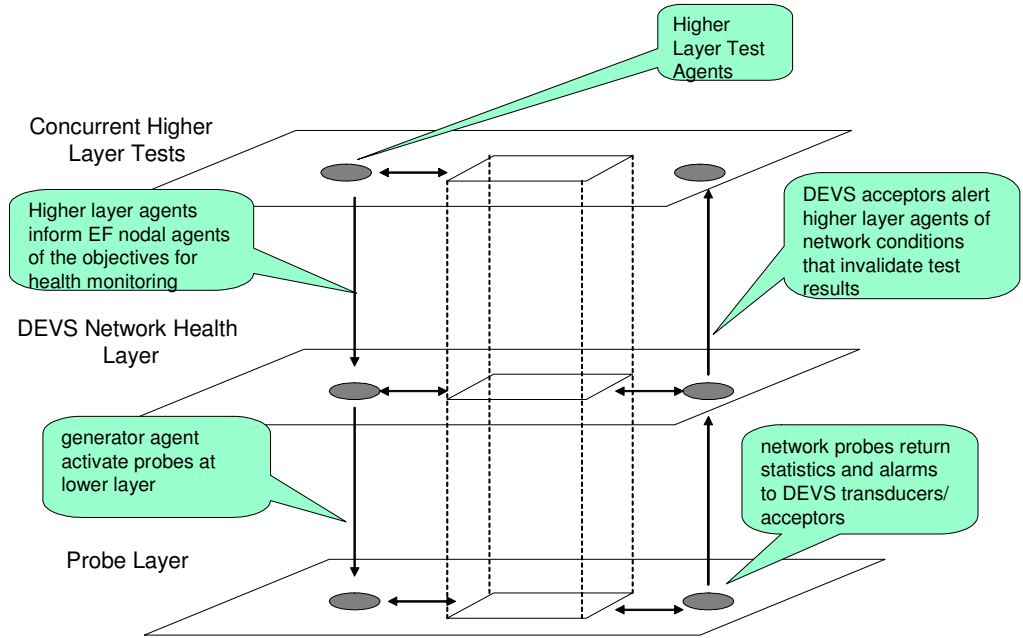

**Figure 34:** Multi-layer testing with Network Health Monitoring

### 7.2.2 Semantic Level – Information Exchange in Collaborations

Mission threads consist of sequences of discrete information exchanges. A collaboration service supports such exchanges by enabling collaborators to employ a variety of media, such as text, audio, and video, in various combinations. For example, a drawing accompanied by a voice explanation involves both graphical and audio media data. Further, the service supports establishing producer/consumer relationships. For example, the graphical/audio combination might be directed to one or more participants interested in that particular item. From a multilevel perspective, testing of such exchanges involves pragmatic, semantic, and syntactic aspects. From the pragmatic point-of-view, the ultimate worth of an exchange is how well it contributes to the successful and timely completion of a mission thread. From the semantic perspective, the measures of performance involve the speed and accuracy with which an information item, such as a graphical/audio combination, is sent from producer to consumer. Accuracy may be measured by comparing the received item to the sent item using appropriate metrics. For example, is the received graphic/audio combination within an acceptable "distance" from the transmitted combination, where distance might be measured by pixel matching in the case of graphics and frequency matching in the case of audio. To automate this kind of comparison, metrics must be chosen that are both discriminative and quick to compute. Further, if translation is involved, the "meaning" of the item must be preserved as discussed above. Also, the delay involved in sending an item from sender to receiver, must be within limits set by human psychology and physiology. Such limits are more stringent where exchanges are contingent on immediately prior ones as in a conversation. Instrumentation of such tests is similar to that at the syntactic level to be discussed next, with the understanding that the complexity of testing for accuracy and speed is of a higher order at the semantic level.

### 7.2.3 Pragmatic Level – Mission Thread Testing

A test federation observes an orchestration of web-services to verify the message flow among participants adheres to information exchange requirements. A mission thread is a series of activities executed by operational nodes and employing the information processing functions of web-services. Test agents watch messages sent and received by the services that host the participating operational nodes. Depending on the mode of testing, the test architecture may, or may not, have knowledge of the driving mission thread under test. If a mission thread is being executed and thread knowledge is available, testing can do a lot more than if it does not.

With knowledge of the thread being executed, DEVS test agents can be aware of the current activity of the operational nodes it is observing. This enables an agent to focus more efficiently on a smaller set of messages that are likely to provide test opportunities.

### 7.2.4 Measuring Success in Mission Thread Executions

The ultimate test of effectiveness of an integration infrastructure is its ability to support successful outcomes of mission thread executions. To measure such effectiveness, the test instrumentation system must be informed about the events and messages to expect during an execution, including those that provide evidence of success or failure, and must be able to detect and track these events and messages throughout the execution.

## *8. Conclusions*

We addressed the problem of net-centricity with the development of DEVS/SOA, which is the SOA implementation of DEVS simulation engine so that models can be executed remotely as well as in a distributed manner using Simulation as a Service within a SOA framework. The DEVS/SOA framework provides the capability to send models to remote locations, run the simulation from other computers and partition the hierarchical coupled model over a set of server farms that host Simulation service.

The integration of enhanced MVC, DEVSML and DEVS/SOA along with the automated model generation from multifarious modes of requirement specifications resulted in a unifying framework called DUNIP (Figure 35).

In this development effort, two implementations of DEVS simulation protocol have been presented. In the first, the simulation process is centralized by means of the Coordinator, which receives and propagates messages from one simulation service to others. There are no changes to the DEVS simulation protocol in this implementation but the real-time Simulation service does require the simulation protocol to be tailored for SOA.

We also described the development of SOA client that provides DEVS-based Services specifically to execute the models as a running simulation. The primary 'simulation' service comprise of many helper services that were also developed. We also went beyond the current SOA framework and proposed a symmetrical SOA that is imperative to distributed execution.

We also demonstrated the DEVS/SOA framework with a real world application of network health monitoring and illustrated the concepts with an example of Joint Close Air Support. This research work has presented proof of concept for DEVS based M&S over SOA. With the enhanced DoDAF [32], automated generation of DEVS model from DoDAF specifications can be executed and the architecture be simulated over a net-centric platform. The DUNIP [31] process also describes many other ways to autogenerate DEVS models from various other types of mission-thread specifications, for example, BPMN/BPEL and message-based restricted Natural Language Progressing (NLP). A Sample demonstration of DUNIP can be seen at [33]. In order to 'execute (as a model)' a set of scenario instructions over net-centric platform, the following capabilities must exist:
1. Transformation of the scenario specifications to a model, which is a DEVS model in this case
2. Execution of model over SOA
3. Communication using XML as middleware.

The first step is described in [31] and step 2 and 3 are presented in this paper. The next stage of analysis of this mission-thread statement is the development of automated test models and their execution over SOA. Automated test-model generation is discussed in [18, 31] and DEVS model execution can be performed by the work presented here.

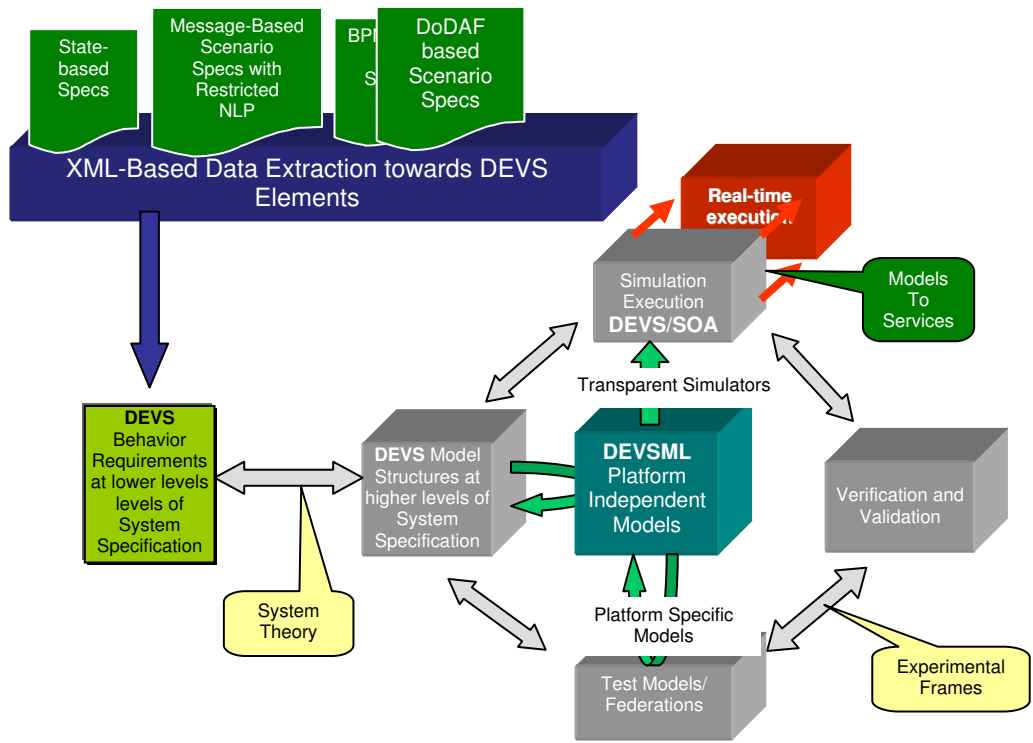

**Figure 35:** The Complete DEVS Unified Process

## 8.1 Future Work

The present research work has the following scope for future development:
- Towards standardization of DEVS formalism [24]
The DEVSML framework developed the atomic and coupled DTDs as meta-models towards collaborative DEVS model development. They are proposed with an idea towards their standardization where the DEVS community can come to a common ground for model reuse and repository management.
- Refine the DUNIP process
A Prototype was demonstrated as a final outcome of this research effort. More features like, validation, consistency checking, etc. should be added to develop it as a COTS product.
- Performance evaluation of distributed DEVS/SOA protocol
The DEVSV/SOA protocol required tailoring of DEVS simulation protocol for SOA domain. Performance evaluation of this version is required to compare it with performance of DEVS protocol with current implementations like DEVS/RMI, DEVS/CORBA etc.
- Make it easier for other DEVS groups to participate in DEVSML and DEVSV/SOA development by registering their simulators
DEVSML is developed as a framework for collaborative model development and portable model specifications resulting from net-centric collaboration using XML middleware. Remote simulation is one capability that is also provided by DEVSML. Various simulator versions from different groups should be gathered and worked upon towards standardized DTDs for an efficient model-sharing system. Currently, two simulator implementations, viz. GenDEVS-ACIMS and xDEVS-Spain have been used to provide proof of concept. Better design of website offering DEVSML service should be designed that would facilitate various groups to submit their simulator implementations.
- Make prototype tool as an Educational aide
The demonstrated prototype should be enhanced for teaching DEVS-based Modeling and Simulation courses. Various manuals and GUI enhancements would be added that facilitate learning and future development.